\documentclass[reprint,aps,amsmath,amssymb,superscriptaddress,nofootinbib]{revtex4-2}
\usepackage{graphicx,bm,amsthm,mathtools, xcolor}
\usepackage[breaklinks=true,colorlinks=false,pdfborder={0 0 0}]{hyperref}
\newcommand{\movie}[1]{\href{https://github.com/R1967J/kinematic-field/raw/main/movies/#1}{\nolinkurl{#1}}}

\theoremstyle{plain}
\newtheorem{theorem}{Theorem}
\newtheorem{proposition}{Proposition}
\newtheorem{lemma}{Lemma}
\theoremstyle{definition}
\newtheorem{definition}{Definition}
\newtheorem{remark}{Remark}

\newcommand{\xv}{\mathbf{x}}\newcommand{\yv}{\mathbf{y}}\newcommand{\uv}{\mathbf{u}}
\newcommand{\pv}{\mathbf{p}}\newcommand{\kv}{\mathbf{k}}\newcommand{\qv}{\mathbf{q}}
\newcommand{\Xv}{\mathbf{X}}\newcommand{\Wv}{\mathbf{W}}
\newcommand{\ev}{\mathbf{e}}\newcommand{\cv}{\mathbf{c}}
\newcommand{\Tb}{\mathbb{T}}\newcommand{\Rb}{\mathbb{R}}\newcommand{\Zb}{\mathbb{Z}}
\newcommand{\Eb}{\mathbb{E}}

\newcommand{\Dsw}{\mathcal{D}}
\allowdisplaybreaks

\begin{document}

\title{Eddies that ride the flow: a model of turbulence with Richardson dispersion and anomalous dissipation}

\author{Raul Jimenez}
\email{raul.jimenez@icc.ub.edu}
\affiliation{Institut de Ci\`encies del Cosmos (ICCUB), Universitat de Barcelona, Mart\'i i Franqu\`es 1, E-08028 Barcelona, Spain}
\affiliation{ICREA, Pg.\ Llu\'is Companys 23, E-08010 Barcelona, Spain}
\author{Svitlana Mayboroda}
\affiliation{Department of Mathematics, ETH Z\"urich, R\"amistrasse 101, 8092 Z\"urich, Switzerland}
\affiliation{School of Mathematics, Institute for Advanced Study, 1 Einstein Drive, Princeton, NJ 08540, USA}
\author{Camillo De Lellis}
\affiliation{School of Mathematics, Institute for Advanced Study, 1 Einstein Drive, Princeton, NJ 08540, USA}
\author{David N. Spergel}
\affiliation{Flatiron Institute, 162 Fifth Avenue, New York, NY 10011, USA}
\author{Max Engelstein}
\affiliation{School of Mathematics, University of Minnesota, Minneapolis, MN 55455, USA}

\date{\today}

\begin{abstract}
We construct a divergence-free velocity field $\uv(\xv,t)$ on $\Tb^{n}$ that is a
genuine function of position and carries both defining properties of turbulent
transport: Richardson pair dispersion and a dissipative anomaly. No synthetic
velocity field has carried both, because the frame that gives the first destroys the
field the second requires. The key
is a dichotomy: deforming a spectrally localised pattern smears it across scales,
whereas pushing its generating centres forward and re-convolving leaves the spectral
envelope exactly invariant --- so transporting the eddy centres preserves both
cascade and frame. In the resulting field the finite-size Lyapunov exponent obeys
$\zeta(R)=h(R)-1$ threshold by threshold within the scaling range, with a sweeping
deficit $\Dsw=+0.038\pm0.019$, consistent with zero, against $+0.091\pm0.019$ for the
frozen field on the same data, a gap that doubles at Navier--Stokes intermittency.  Companion papers provide rigorous proof that this family of solutions shows Richardson dispersion and dissipative anomaly \cite{FeffermanMayboroda2026}, and that a scalar field advected by this flow obeys Yaglom's relation \cite{delellis}, for small Kubo number.  Trajectories in it forget their starting point as the smallest eddy is removed: the
median separation of pairs released from one point with independent Brownian kicks
varies as $\kappa_{p}^{0.0333\pm0.0038}$ over twenty-two decades of particle
diffusivity, against $\kappa_{p}^{1/2}$ for a smooth control. A scalar in the same field closes the
exact scale-by-scale variance budget, four independently measured terms balancing the
dissipation to $15\%$ over $1.2$ decades with no adjustable constant, and reaches its
inertial--convective asymptote
$S_{L}=-\chi r$ to $0.9995\pm0.0018$ once the measured diffusive transfer
is restored, while under an imposed mean gradient
$\chi$ changes by $14\%$ over a factor $5.8$ in $\kappa$.
\end{abstract}
\maketitle

\section{Introduction}

Two facts organise the phenomenology of high-Reynolds-number turbulence. The first
is Richardson's: two fluid particles at separation $R$ inside the inertial range
separate at a rate set by the eddies of size $R$ and by nothing else, so that
$d\langle R^{2}\rangle/dt\propto\varepsilon^{1/3}\langle R^{2}\rangle^{2/3}$ and
$\langle R^{2}\rangle\sim\varepsilon t^{3}$ \cite{Richardson1926,Obukhov1941,%
Batchelor1950}. The second is Onsager's: the rate at which a turbulent flow
destroys a conserved quantity does not vanish as the corresponding molecular
coefficient is taken to zero \cite{Onsager1949,EyinkSreenivasan2006}. For a
passive scalar $\phi$ this is the statement that
$\chi=2\kappa\langle|\nabla\phi|^{2}\rangle$ tends to a positive limit as
$\kappa\to0$ \cite{Obukhov1949,Corrsin1951,Donzis2005}. For the kinetic energy this phenomenon is related to the regularity of the
velocity field. If the field is smoother than H\"older $1/3$, dissipation is
impossible \cite{ConstantinETiti1994}, whereas below that threshold it can occur
\cite{Isett2018,DLS2019}.

The two facts are not independent. For a passive scalar advected by a divergence-free
field, away from walls, anomalous dissipation and Lagrangian spontaneous
stochasticity are equivalent \cite{DrivasEyink2017}: the scalar can only lose
variance in the vanishing-diffusivity limit if trajectories emanating from a single
point remain random in that limit, which is Richardson's explosive separation
carried to $R\to0$. A model that produced one without the other would be
suspect; a vector field that produced both, with independently
prescribed regularity, would be a laboratory in which the equivalence could be
examined directly.

The ability to construct such a vector field would address a gap left by the two current approaches.
Direct simulation of the Navier--Stokes equations resolves the dynamics but not the
asymptotics: the scalar anomaly requires
$R_{\lambda}\mathrm{Sc}^{1/2}\gg1$ at $\mathrm{Sc}<1$ and
$\ln(\mathrm{Sc})/R_{\lambda}\ll1$ at $\mathrm{Sc}>1$ \cite{Donzis2005}, and a clean $t^{3}$ range for
pair separation has arguably never been observed at all, with the accessible
window dominated by the ballistic Batchelor regime
\cite{Bourgoin2006,Elsinga2022}; the reviews
\cite{Sawford2001,SalazarCollins2009} survey what has and has not been seen. At the other extreme, the rigorous programme has
in the last few years produced velocity fields for which anomalous dissipation is a
theorem --- H\"older-continuous drifts built by iterated homogenisation
\cite{ArmstrongVicol2025}, via mixing \cite{ColomboCrippaSorella2023},
and, most recently, drifts that are themselves weak solutions of the Euler
equations \cite{Burczak2023}. Richardson's law itself has been proved for an explicitly
engineered deterministic field \cite{HessChildsRowan2025}. These constructions are engineered to make a proof
close, but they are not physical, and their Lagrangian statistics beyond the one each was built
for are largely unexplored. Only one of them
has been argued to carry the sweeping structure of a physical flow at all
\cite{Eyink2024,ArmstrongVicol2025}, but its time structure is unrealistically
lacunary.

Between the two sits the synthetic, or \emph{kinematic}, velocity field built as a random
superposition of eddies with prescribed amplitudes, lifetimes and scales, in which
the inertial range is as wide as one chooses to make it and the H\"older exponent
is an input rather than an output. Two constructions carry the same name and should be kept apart. Kinematic
simulation (KS) in the narrow sense is a sum of random \emph{Fourier} modes with a
prescribed spectrum and a prescribed mode frequency
\cite{Kraichnan1970,Fung1992}, and it is that method whose Lagrangian statistics have
been studied at length \cite{FungVassilicos1998,MalikVassilicos1999}. What is used
below is instead a superposition of \emph{localised} eddies, and its skeleton is not
new: \c{C}a\u{g}lar \cite{Caglar2007}, developing a construction of \c{C}inlar
\cite{CaglarCinlar2001}, writes a fixed divergence-free profile convolved with a
Poisson measure of centres over a continuum of scales, and chooses the arrival law so
that $E(|\kv|)\propto|\kv|^{-5/3}$ across an inertial range. There, as in the
vortex-superposition models it descends from
\cite{Townsend1976,Lundgren1982,PullinSaffman1993}, the centres are placed and left
where they are placed. This paper is about what happens when they are not. It is the
natural laboratory. It has also, for the whole of its history, failed at Richardson's
law.
\subsection{What Richardson's law asks of a velocity field}
\label{sec:asks}

The failure has a precise anatomy. Richardson's law rests on three
requirements, and they are worth separating because the synthetic field
supplies the first two by construction and has, until now, failed at the
third. Consider a pair at separation $r$ inside the inertial range, let
$\delta u(r)$ be the typical velocity difference across $r$, and let
$\tau(r)$ be the time over which that difference stays correlated along
the pair. The pair diffusivity is then $K(r)\sim\delta u(r)^2\,\tau(r)$,
and $d\langle r^2\rangle/dt\sim K(r)$.

\emph{Scaling.} The increment obeys $\delta u(r)\sim(\varepsilon r)^{1/3}$,
which is Obukhov's $k^{-5/3}$ spectrum read in physical space \cite{Obukhov1941}. More
generally $\delta u(r)\sim r^{h}$, with $h$ the H\"older exponent of the
field, and nothing below depends on $h=1/3$.

\emph{Clock.} An eddy of size $r$ turns over in $r/\delta u(r)$ and lives
for a fixed number of turnover times,
\begin{equation}
\tau_{\rm eddy}(r)=\mathrm{Ku}\,\frac{r}{\delta u(r)}
\sim\mathrm{Ku}\,\varepsilon^{-1/3}r^{2/3},
\label{eq:clock}
\end{equation}
where the Kubo number Ku, the ratio of the lifetime to the turnover time,
is of order one in a real flow.

\emph{Advection.} The pair must stay with the eddies of its own size for
their lifetime. The pair is carried by every scale coarser than $r$; for
the eddies of size $r$ to keep it, those scales must carry the eddies as
well, so that the sweep is common to both and cancels from their relative
motion. Then $\tau(r)=\tau_{\rm eddy}(r)$ and
\begin{equation}
K(r)\sim\mathrm{Ku}\,\varepsilon^{1/3}r^{4/3},\qquad
\langle r^2\rangle\sim\mathrm{Ku}^3\,\varepsilon\,t^3,
\label{eq:richardson}
\end{equation}
Richardson's four-thirds law and its integral \cite{Richardson1926,Batchelor1950}; the factor
$\mathrm{Ku}^3$ is the one in the lower bound of Ref.~\cite{FeffermanMayboroda2026}. Read at fixed
scale rather than at fixed time, the same three ingredients give a
separation rate $\lambda(r)\sim1/\tau_{\rm eddy}(r)\sim r^{h-1}$, which is
the form in which the law is tested in Section~\ref{sec:richardson}.

The first two requirements are properties of the field at a point and can
be prescribed; in the model below they are the input exponent $\alpha$ and
the lifetime $\tau_j=\mathrm{Ku}\,\ell_j/\bar a_j$ of Eq.~\eqref{eq:scales}. The third is
a property of the field along a pair, and it is where the synthetic field
fails. The cause was identified by Thomson and Devenish \cite{ThomsonDevenish2005}. In a real
flow the energetic large eddies advect both the particles and the small
eddies, so the two are swept together. In a synthetic field whose eddies
are placed and left where they are placed, the large modes advect the
particles only: the small-scale pattern stands still while the pair is
dragged through it (Fig.~\ref{fig:one}A), and the pair loses the eddies of its own size
not after $\tau_{\rm eddy}(r)$ but after the far shorter sweeping time
$r/U_0$, where $U_0$ is the r.m.s.\ speed. Eyink and Benveniste \cite{EyinkBenveniste2013}
sharpened this into a formula: the pair diffusivity is
$K_{ij}(r,t)=S_{ij}(r)\,\tau(r,t)$ with
\begin{equation}
\tau(r,t)=\min\{\tau_{\rm eddy}(r),\,t,\,r/U_0\},
\label{eq:EB}
\end{equation}
and the third entry is the one that has no counterpart in Navier--Stokes.
Substituting it gives $K\sim\varepsilon^{2/3}r^{5/3}/U_0$ and
$\langle r^2\rangle\sim\varepsilon^4t^6/U_0^6$; in the absence of a mean
flow the regions of anomalously small velocity contribute a $t^3$
population on top of a $t^6$ bulk and the mean grows as $t^{9/2}$ \cite{ThomsonDevenish2005}.

Whether the effect is fatal in practice has been contested. Osborne
\emph{et al.} \cite{Osborne2006} reported Richardson scaling in three-dimensional KS
despite the absence of sweeping and attributed it to the fractal set of
straining stagnation points, for which D\'avila and Vassilicos had earlier
derived $\gamma=6/D_s$ with $D_s=2$ \cite{DavilaVassilicos2003}; Devenish and Thomson \cite{DevenishThomson2009}
answered that the discrepancy was not a time-stepping artefact; Malik \cite{Malik2017}
argued that the sweeping error in the pair diffusivity decreases with
separation and so cannot explain the observed non-locality.
Two-dimensional KS reproduces the measured separation distributions of
laboratory and simulated two-dimensional turbulence \cite{NicolleauVassilicos2003,Jullien1999} while showing
an initial-condition dependence of the r.m.s.\ separation that persists
for very long times --- which is precisely the quantity
Eq.~\eqref{eq:EB} attacks. Our position is the conservative one: rather
than adjudicate the debate, we remove the third entry of
Eq.~\eqref{eq:EB} by construction, and verify against a control that
shares every random number with the experiment.

To Richardson's three requirements the anomaly adds a fourth, of a
different kind. The dissipation $\chi=2\kappa\langle|\nabla\phi|^2\rangle$
is defined by an advection--diffusion equation, and an
advection--diffusion equation needs a velocity field $u(x,t)$: a rule
that moves particles but assigns different velocities to the same point
at the same time will not do. The third requirement is Lagrangian and the
fourth is Eulerian, and the obvious way to meet the third defeats the
fourth (Section~\ref{sec:framefield}). In the Euler equations there is no such conflict.
The velocity is a field, and every structure in it, at every scale, is
carried by the flow, so that the third requirement is not an assumption
but the content of Kelvin's theorem. To see what a model must keep and
what it may replace, we write Euler in the formulation that makes this
transport explicit.

\subsection{Euler in Lagrangian labels}
\label{sec:lagrangian}

Let $X(a,t)$ be the flow map of a solution of the incompressible Euler
equations on $\mathbb T^n$ and let $A(x,t)=X(\cdot,t)^{-1}(x)$ be the
back-to-labels map (Constantin's letter; the amplitudes $A_j$ of Section
II are scalars and no confusion should arise). It is transported by the
flow,
\begin{equation}
\partial_tA+u\cdot\nabla A=0,\qquad A(x,0)=x,\qquad\det\nabla A=1,
\label{eq:labels}
\end{equation}
and it carries all Lagrangian information, since $\nabla A=F^{-1}\circ A$
with $F=\nabla_aX$. Kelvin's theorem states that the velocity 1-form is
transported up to an exact form; pulled back to Eulerian coordinates this
is the Weber formula
\begin{equation}
u=\mathbb P\big[(\nabla A)^{\!\top}u_0\circ A\big],
\label{eq:weber}
\end{equation}
with $\mathbb P$ the Leray projection, and its curl is Cauchy's formula
\begin{equation}
\omega=(\nabla A)^{-1}\,\omega_0\circ A\quad(n=3),\qquad
\omega=\omega_0\circ A\quad(n=2).
\label{eq:cauchy}
\end{equation}
Either formula closes the system on $A$ alone: a first-order transport
equation whose velocity depends nonlocally on $\nabla A$ through a
Calder\'on--Zygmund operator. The pressure has disappeared, and the only
evolving unknown is the label map. Constantin proved local well-posedness
in this framework and used it for the vanishing-viscosity limit \cite{Constantin2001local,Constantin2001NS};
the same Cauchy invariants underlie the time-analyticity of Lagrangian
trajectories \cite{FrischZheligovsky2014,ConstantinVicolWu2015}. In Lagrangian variables the content of
Eq.~\eqref{eq:cauchy} is $\partial_tF=(\nabla u\circ X)\,F$, and with the
decomposition $\nabla u=S+\tfrac12\,\omega\times$, where $S$ is the
symmetric strain, a singular integral of $\omega$, the rotation drops out
of the vorticity equation:
\begin{equation}
\frac{d}{dt}\,\omega\circ X=(S\circ X)\,\omega\circ X.
\label{eq:stretch}
\end{equation}

Now expand the initial vorticity in divergence-free wavelets,
$\omega_0=\sum_\lambda c_\lambda\Omega_\lambda$, $\lambda=(j,i)$, with
$\Omega_\lambda$ supported in the ball $B_\lambda=B(y^j_i,R\ell_j)$.
Taking $\Omega_\lambda=\nabla\times W_j(\cdot-y^j_i)$, with $W_j$ the
divergence-free model eddy of Section~\ref{sec:basis}, makes the velocity of each
element at $t=0$ equal to the eddy itself. Since the coefficients are
Cauchy invariants,
\begin{equation}
\omega(x,t)=\sum_\lambda c_\lambda\,(\nabla A)^{-1}\Omega_\lambda
\big(A(x,t)\big),\qquad u=K*\omega,
\label{eq:elements}
\end{equation}
so for all time the vorticity is a superposition of the same elements,
each carried by the flow. The element $\lambda$ lives on $X(B_\lambda,t)$,
its vorticity vector is $F\Omega_\lambda$, and it induces on every other
element the Biot--Savart field of its deformed self. Its centre
$X_\lambda=X(y^j_i,t)$ and its shape obey
\begin{equation}
\dot X_\lambda=u(X_\lambda,t),\qquad
\dot F_\lambda=\nabla u(X_\lambda,t)\,F_\lambda,
\label{eq:centre-shape}
\end{equation}
and to first order in $\ell_j$ the support is the ellipsoid
$X_\lambda+F_\lambda B(0,R\ell_j)$, with semi-axes $R\ell_j\sigma_k$ and
$\prod_k\sigma_k=1$.

What deforms an element of size $\ell_j$ is the velocity increment across
its support. Under Kolmogorov scaling that increment receives
$\ell_j\ell_m^{-\beta}$ from each coarser scale $m<j$ and $\ell_m^{\alpha}$
from each finer scale $m>j$, where $\beta=1-\alpha$; both geometric sums
are dominated by $m\approx j$ and total $\sim\ell_j^{\alpha}$. Hence the
strain rate felt by the element is $\ell_j^{\alpha-1}=\ell_j^{-\beta}$:
straining is local in scale, and the element is deformed by $O(1)$ after
one turnover time $\ell_j^{\beta}$. Heuristically, beyond that time
$F_\lambda$ grows exponentially; the element is stretched in one or two
directions and, since $\det F_\lambda=1$, thinned in the others, so that
its Eulerian wavelet content migrates to scales
$\ell_j/\sigma_{\max}\ll\ell_j$. This is the cascade seen from the labels,
and it has two consequences for the Eulerian statistics at scale $\ell_j$:
the energy there is supplied by the thinning of coarser elements, and the
field decorrelates on the turnover time, with a Kubo number of order one.

This is how Euler meets the four requirements of Section~\ref{sec:asks}, and it says
what each costs. The first, $\delta u(\ell_j)\sim\ell_j^{\alpha}$, is the
Kolmogorov--Obukhov phenomenology, an input to the argument just given and
not a theorem. The second and the third are the two halves of
Eq.~\eqref{eq:centre-shape}: the centre is carried by the full velocity,
so the pair and the elements of its own size are swept together and what
separates the pair is the increment across those elements; and the shape
stays coherent for one turnover time, so $\mathrm{Ku}=O(1)$ is an output
and not a parameter. The fourth is automatic, since $u=K*\omega$. The
price is that none of this is rigorous: $S$ is a singular integral over
all elements, and in the limit of infinitely many scales the pointwise $F$
is not even defined, since the flow map of a H\"older field is at best
H\"older. What a model must keep is now visible: elements that are
carried, as in the first half of Eq.~\eqref{eq:centre-shape}, and that
live for a time set by their own scale. What it may replace is the second
half, the deformation, provided it replaces what the deformation does.

\subsection{The renewal model as a closure}
\label{sec:renewal}

The model of \cite{FeffermanMayboroda2026,delellis} and of this paper keeps the elements and the
sweeping and replaces everything else by a stochastic rule. Its field is
$u=\sum_ju_j$ with
\begin{eqnarray}
u_j(x,t)&=&A_j\sum_{k}g_{j,k}(t)\sum_i\eta^{(j,k)}_i
W_j\big(x-p^{(j,k)}_i(t)\big) \nonumber\\
\dot p^{(j,k)}_i&=&u_{<j}\big(p^{(j,k)}_i,t\big),
\label{eq:model}
\end{eqnarray}
where $u_{<j}=\sum_{m<j}u_m$, the epoch $(j,k)$ is a window of length of
order $\tau_j=\mathrm{Ku}\,\ell_j/\bar a_j$ --- the clock of
Eq.~\eqref{eq:clock} --- and the independent, mean-zero weights $\eta$ and
the uniformly shifted birth lattice of the centres are drawn afresh for
every epoch; the precise forms are Eqs.~\eqref{eq:scales}--\eqref{eq:centers-flow} of
Sections~\ref{sec:basis} and \ref{sec:comotion}.
Compared with the Euler element of
Eqs.~\eqref{eq:elements}--\eqref{eq:centre-shape}:

(i) The centre moves with $u_{<j}(p)$ instead of $u(X_\lambda)$: sweeping
by the coarser scales is kept exactly, while self-induced motion and
advection by the finer scales are dropped. This is what makes the coupling
of the centres to the field triangular rather than circular, so that the
construction closes generation by generation, and what makes the weights
of generation $j$ independent of the positions of its centres, which
Theorem~\ref{thm:band} uses.

(ii) $F_\lambda\equiv I$: no strain, no rotation by $\tfrac12\omega_{<j}$,
no stretching. Orientation and amplitude $A_j$ are pinned, so Kolmogorov
scaling is an input rather than an output, and the stationarity of the
spectrum, which Euler would have to produce by transfer, is imposed.

(iii) Each element contributes the compactly supported $W_j$ itself rather
than the Biot--Savart field of its deformed vorticity: there is no
cross-scale induced velocity and no pressure.

(iv) The refresh. Over one window the deformation that Euler would apply
is small, $\|F_{<j}-I\|\le e^{C\,\mathrm{Ku}}-1$ (Lemma 2.2 of
Ref.~\cite{FeffermanMayboroda2026}), and it is not wholly discarded: the particles follow the full
$u$ at their own positions, so the mismatch between the strained pair and
the unstrained eddy is the commutator
\begin{equation}
B_au_b=A_b\sum_i\eta_i\big[u_a(y)-u_a(p^b_i)\big]\cdot
\nabla W_b\big(y-p^b_i\big),\qquad a<b,
\label{eq:commutator}
\end{equation}
which is the linearised deformation. It is kept in the proof as an error
of relative size $\mathrm{Ku}\,(\ell_b/\ell_a)^{\beta}$ per window (see
the proof of Lemma 4.3 of Ref.~\cite{FeffermanMayboroda2026}) and is summable over scales. In the
field of this paper it is the residual sweeping
$e_i=u_{<j}(X)-u_{<j}(p_i)$ of Proposition~\ref{prop:residual}, bounded on the support of
the eddy by a quantity of the order of the eddy's own velocity
(Eq.~\eqref{eq:errsize}).

The model thus applies the first-order deformation for a time $\tau_j$, as
seen by the pair, and then discards the element and redraws it. The
refresh is a closure for the three effects that the $O(1)$ deformation
would produce beyond a turnover time: the distortion of the element, its
transfer to finer scales, and the decorrelation of scale $\ell_j$. All
three are replaced by a single stochastic step with prescribed statistics,
and this step is what makes the proof work: averaging over the fresh data
annihilates the first-order and the mixed terms exactly (Prop.~6.6 of
Ref.~\cite{FeffermanMayboroda2026}), leaving the gain $\tau_j^2\Gamma_j$, whereas in Euler the same
cancellation would have to come from the mixing of the label map, which is
not available.

Two features of the rule have no counterpart in Euler and are the price of
this closure. The Kubo number must be small: the element is refreshed at
time $\mathrm{Ku}\,\ell_j^{\beta}$, well before Euler's natural coherence
time $\ell_j^{\beta}$, which is precisely the regime in which rigid
translation followed by refresh is self-consistent, whereas real
turbulence has $\mathrm{Ku}=O(1)$. And the refresh times are nested
dyadically, which the averaging and the time reversal in the dissipation
argument require. The authors of Ref.~\cite{FeffermanMayboroda2026} regard the smallness of Ku as
technical; a derivation of the model from Euler would amount to showing
that the deformation of Lagrangian wavelet elements over a turnover time
produces statistics that approximate an independent, mean-zero refresh.

The field of Definition~\ref{def:main} is this model carried into the regime the proof
does not reach, and it differs from the proved version in the following
particulars, each chosen so that something can be measured. The abrupt
refresh is replaced by the $\cos^2$ cross-fade of Eq.~\eqref{eq:window}, under which the
epochs form an exact partition of unity: $u$ is continuous in $t$, nothing
is paused across a renewal, and the identity of Theorem~\ref{thm:band} holds at every
instant. The bounded symmetric weights are replaced by the log-normal
weights of Eq.~\eqref{eq:weights}, whose parameter $c_\mu$ sets the intermittency of the
field from none to that of Navier--Stokes turbulence and beyond, so that
the Richardson test can be repeated as a function of intermittency
(Fig.~\ref{fig:four}C). The Kubo number is of order one, from 0.25 to 2, the regime of
real turbulence: there the residual of Eq.~\eqref{eq:commutator} is of
order $A_j$ and is no longer controlled by the proof, and whether it
decorrelates on the eddy's own clock or on the far shorter clock of the
coarse field that produces it becomes a quantitative question, which the
measurements of Section~\ref{sec:richardson} settle. The scale ratio $s$ is free in $(1,2]$:
the half-octave bands used for the dispersion runs sample the same
spectrum twice as finely, and they, not the octaves, produce a plateau in
the realised $h(r)$ (Fig.~\ref{fig:two}A); the epochs are then not nested across
generations. The profile is a $\cos^2$ bump rather than the radial $C^4$
stream function of the proof, and the one property of the profile the
proof needs, its condition (ND), is checked for the bump direction by
direction (Appendix~\ref{app:methods}). In both versions the energy of each generation is
fixed by construction and no equation moves energy between scales: the
field carries no cascade, and the diagnostics of Section~\ref{sec:nocascade} confirm
it.

\begin{figure*}[t]
\centering
\includegraphics[width=\textwidth]{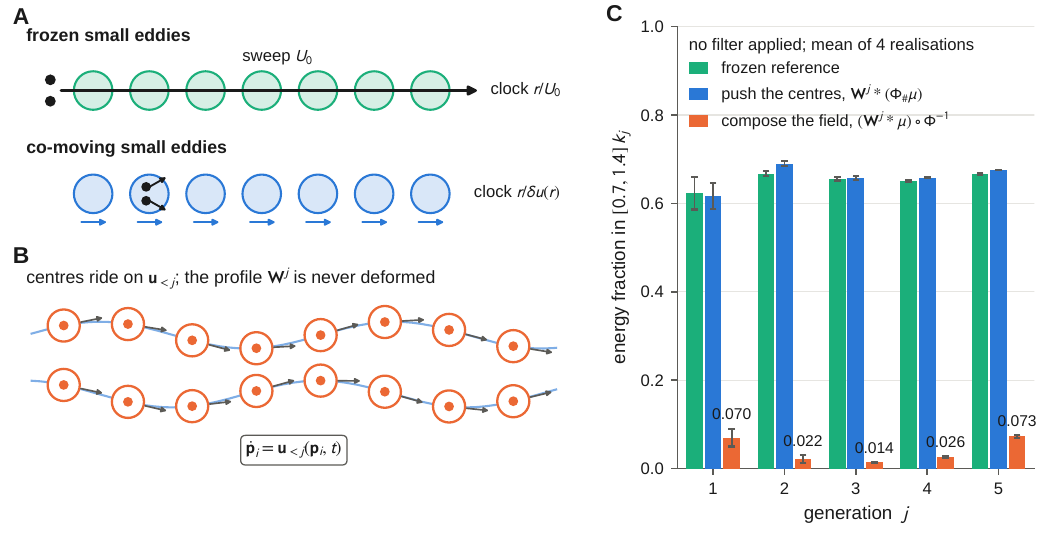}
\caption{\textbf{From the sweeping obstruction to a single-valued Eulerian field.}
(\emph{A}) In a frozen multiscale field the energetic large modes advect the pair
but not the small eddies, so the pair is dragged through a stationary pattern and
decorrelates on $r/U_{0}$; when the small scales are carried by the same coarse
flow the sweep is common to both and cancels from the relative motion, restoring
the turnover clock $r/\delta u(r)$. (\emph{B}) The construction: the centres of
generation $j$ are advected by $\uv_{<j}$, the sum of all coarser generations
(Eq.~\eqref{eq:centres}), while the profile $\Wv^{j}$ is translated rigidly and
never deformed (Eq.~\eqref{eq:field}). (\emph{C}) Verification of
Theorem~\ref{thm:band}: fraction of generation $j$'s energy remaining inside its
own octave band, averaged over four realisations (bars: standard error), with no
filter applied. Pushing the generating centres forward reproduces the frozen reference to
within $0.022$ in absolute fraction; composing
the field with the same flow map destroys the cascade.}
\label{fig:one}
\end{figure*}

Item (ii) deserves a second look, because it is where the construction is
most easily spoiled, and it is the content of this paper. In Euler the
deformation $F_\lambda$ \emph{is} the cascade --- it moves the element's
wavelet content to finer scales --- and the transfer term of the energy
equation is what replenishes each scale. A model that prescribes the
energy of each generation has no such term, so if it deforms its elements
it loses them from their band and nothing replaces them. The alternative
can be stated exactly, and is Lemma~\ref{lem:dichotomy} below. A spectrally localised
pattern is not an arbitrary function but a convolution $W_j*\mu$ of a
fixed profile with an atomic measure of centres, and there are two
inequivalent ways to move it. Composing it with the flow map,
$(W_j*\mu)\circ\Phi^{-1}$, is Euler's transport with the true $F_\lambda$:
it multiplies the pattern's gradient by the carrier's strain and smears
its spectrum across wavenumbers, and on the field of this paper it leaves
only 1.4 to 7.3\% of each generation's energy in its band (Fig.~\ref{fig:one}C).
Pushing the measure forward and re-convolving, $W_j*(\Phi_\#\mu)$, is
$F_\lambda\equiv I$: it changes only the phases of $\hat\mu$, and the
expected spectral envelope $\widehat W_j$ is exactly invariant, at every
time and along every trajectory the centres take (Theorem~\ref{thm:band}). The two
agree whenever $\nabla\Phi$ is a constant orthogonal matrix preserving the
profile and part company as soon as $\nabla\Phi$ leaves $O(n)$. The
distinction is invisible at the level of a translation, however large,
which is presumably why it has not been isolated before; it is the
gradient of the displacement --- the strain of the carrier, the thing one
is trying to model --- that separates them. Transporting the eddy centres,
each generation carried by the sum of all coarser generations, therefore
keeps the cascade intact while realising the quasi-Lagrangian cancellation
up to the residual of Eq.~\eqref{eq:commutator}.

\subsection{The frame, the field, and what this paper does}
\label{sec:framefield}

The renewal model is one closure of Eq.~\eqref{eq:centre-shape}, and the
alternatives are instructive, because each keeps something it gives up.
The oldest removes the sweep from the observer rather than from the field.
Belinicher and L'vov \cite{BelinicherLvov1987} observe the field from a frame riding with a
marked fluid particle, $v(t,x)=u\big(t,x-\varrho(t)\big)$ with
$\dot\varrho(t)=u\big(t,x_0+\varrho(t)\big)$; because the sweeping
velocity is common to the observer and to the small scales, it cancels
from velocity increments and the infrared divergences it generates
disappear order by order. Applied to a kinematic model the recipe is to
give each particle $a$ its own displacement registers
$D^{(m)}_a(t)=\int_0^tu^{(m)}_a$, one per generation, to read octave $j$
at the shifted point $s^{(j)}_a=X_a-\sum_{m<j}D^{(m)}_a$, and to advance
$\dot X_a=\sum_ju^{(j)}_a$. Then
\begin{equation}
\frac{d}{dt}\,s^{(j)}_a=\sum_{m\ge j}u^{(m)}_a,
\label{eq:frame}
\end{equation}
so the point at which octave $j$ is read moves under octaves $\ge j$ only,
and every coarser octave is removed from the relative motion identically.
This is what Lacorata and Vulpiani \cite{LacorataVulpiani2017} implement, and it yields the
desired statistics: with quasi-Lagrangian coordinates a chaotic multiscale
model recovers the expected finite-size Lyapunov scaling.

In the language of Section~\ref{sec:lagrangian} the registers are a label map built one
trajectory at a time: $\sum_{m<j}D^{(m)}_a$ stands in for $x-A(x,t)$ at
the scales coarser than $j$, but it is a functional of the whole
trajectory of $a$ and not a function of $x$, and that is what it costs.
Two particles occupying the same point at the same time with different
histories are assigned different velocities. What Eq.~\eqref{eq:frame}
defines is a map $\mathcal V:\mathbb T^n\times(\mathbb R^n)^J\to\mathbb
R^n$ on an extended phase space; the object it induces on $\mathbb T^n$ is
the relation $\{(x,\mathcal V(x,D)):D\in(\mathbb R^n)^J\}$, which is not a
graph. There is no $u(x,t)$ with $\dot X=u(X,t)$, and therefore no
divergence, no Jacobian, and no advection--diffusion equation: the third
requirement of Section~\ref{sec:asks} is met exactly and the fourth is lost. In the
model studied below, the fraction of each octave's kinetic energy that is
not a function of position is 0.0003, 0.716, 0.938, 0.972, 0.973, 0.993
for $j=0,\dots,5$: only the largest octave, whose shift is empty, is
single-valued. The renewal model makes the opposite exchange. Its
centres, finitely many of them, solve ordinary differential equations in
$t$ alone, so the field is a function of position, and the frame holds
only up to the residual of Proposition~\ref{prop:residual}, of order $A_j$. Exactness of
the frame for existence of the field is the substance of the paper.

Two constructions do transport localised structure in a field, and in each
the element is deformed. Ghate and Lele \cite{GhateLele2017,GhateLele2020} enrich a large-eddy
simulation with localised Gabor modes and advect their centres,
$dx_i/dt=U_i$, explicitly in order to put the small scales in the sweeping
frame. But the carrier is a separately resolved LES field, coupled one
way, and the enrichment spans a single band of a factor four to eight
rather than a cascade; the wavevectors refract along the trajectory,
$dk_i/dt=-k_m\partial_iU_m$, which is the deformation $F_\lambda$ of
Eq.~\eqref{eq:centre-shape} acting on the wavevector, so the profile does
not stay fixed and the envelope is not preserved; and the assembled field
is divergence-free only to leading order in the window width. Armstrong
and Vicol \cite{ArmstrongVicol2025} do carry every scale in the frame of the coarser ones, but
by composition --- the true $F_\lambda$, route (ii) of Lemma~\ref{lem:dichotomy} --- and
their inverse flow maps must then be renewed on a time scale short
compared with the inverse Lipschitz constant, with the shears paused
across each renewal. Neither reports pair dispersion, and neither reports
a scalar anomaly.

So the two goals have been mutually exclusive. Richardson's law needs the
frame; the anomaly needs a field; and the frame is not a field. We
construct a field that is both. The construction occupies
Sections~\ref{sec:basis}--\ref{sec:euler}: the generating data --- scales,
profile, epochs and weights --- in Section~\ref{sec:basis}, the co-motion of the
centres in Section~\ref{sec:comotion}, and in Section~\ref{sec:euler} the
Eulerisation by push-forward, with Lemma~\ref{lem:dichotomy} and
Definition~\ref{def:main}. Section~\ref{sec:props} collects the properties of the result: it is a function of position and
exactly divergence-free; its flow preserves measure and orientation and
does not fold (Proposition~\ref{prop:flow}); the transport leaves the expected spectral
envelope of every generation unchanged (Theorem~\ref{thm:band}); the residual sweeping
is bounded by a quantity of the order of the eddy's own velocity
(Proposition~\ref{prop:residual}); and the field converges in $L^2$ as generations
are added (Theorem~\ref{thm:L2-limit}). Section~\ref{sec:richardson} puts pairs in
it, against a frozen control sharing
every random number. The finite-size Lyapunov exponent of the co-moving
field obeys $\zeta(R)=h(R)-1$ threshold by threshold within the scaling
range, with a deficit $\mathcal D=+0.038\pm0.019$, consistent with zero,
against $+0.091\pm0.019$ for the frozen field; the paired gap,
$+0.052\pm0.014$ ($3.8\sigma$), doubles at the intermittency of
Navier--Stokes turbulence, persists at Kubo numbers from 0.25 to 2 and at
twice the resolution, and exceeds by a factor of sixty-six the effect of
replacing the birth lattice by a Poisson process. The same section tests
spontaneous stochasticity directly: pairs released from one point with
independent Brownian kicks have a median separation that varies as
$\kappa_p^{0.0333\pm0.0038}$ over twenty-two decades of particle
diffusivity, against $\kappa_p^{1/2}$ for a smooth control, with a
residual that falls as $\ell_{\min}^{0.47\pm0.05}$ as the cutoff is
removed and is the same forward and backward in time. Section~\ref{sec:anomaly} places a
scalar in the field, which only a field permits: the exact scale-by-scale
budget closes, four independently measured terms balancing the dissipation
to 15\% over 1.2 decades with no adjustable constant; the
inertial--convective asymptote $S_L=-\chi r$ is reached to
$0.9995\pm0.0018$ once the measured diffusive transfer is restored; and
under a mean gradient $d\log\chi/d\log\kappa=-0.088\pm0.010$, where a
scalar without an anomaly gives $+1$. Section~\ref{sec:discussion} discusses what has been
constructed, the part played by intermittency, the absence of an energy
cascade and the origin of the arrow of time of the anomaly, and the
relation to the rigorous constructions.

What is new is therefore not the superposition but its motion, in four
respects. The dichotomy of Lemma~\ref{lem:dichotomy} --- that a spectrally localised pattern
admits two inequivalent transports, agreeing only where $\nabla\Phi$ is
orthogonal --- is what makes the choice a choice, and we have not found it
stated. Once it is made, Theorem~\ref{thm:band} is an identity in $k$, at every time
and along every trajectory the centres take, with no hypothesis on their
arrangement: nothing is renewed and nothing degrades. The carrier is
internal, generation $j$ being advected by the sum of the coarser members
of the same construction, so the recursion closes at any depth and no
resolved outer field enters. And the multivaluedness of the
quasi-Lagrangian alternative is measured rather than asserted, octave by
octave. To these the derivation of Sections~\ref{sec:lagrangian} and \ref{sec:renewal} adds a
placement: the model is the Lagrangian wavelet decomposition of the Euler
equations with one closure, the refresh of the elements in place of their
deformation, whose price in the proof --- a small Kubo number --- is here
examined at Kubo numbers of order one. What those buy is the pair of
measurements that follow, which no synthetic velocity field has carried at
once.

\section{The generating data}
\label{sec:basis}

We work on the torus  $\Tb^{n}=(\Rb/L\Zb)^{n}$; lengths are in grid cells and
$L=N$.

\paragraph{Scales.} Fix an integral scale $\ell_{0}$, a H\"older exponent
$\alpha\in(0,1)$, a Kubo number $\mathrm{Ku}$, and a scale ratio $s\in(1,2]$. For
$j=0,\dots,J-1$ put
\begin{equation}
\ell_{j}=\ell_{0}s^{-j},\quad
\bar a_{j}=s^{-j\alpha},\quad
A_{j}=\bar a_{j}\sqrt{\tfrac{\ln s}{\ln 2}},\quad
\tau_{j}=\mathrm{Ku}\,\frac{\ell_{j}}{\bar a_{j}} .
\label{eq:scales}
\end{equation}
Thus $\ell_{j}$ is the spatial scale of generation $j$, $A_{j}$ its velocity
amplitude, and $\tau_{j}=\mathrm{Ku}\,\ell_{0}s^{-j(1-\alpha)}$ its lifetime --- the
turnover time of an eddy of size $\ell_{j}$ moving at the bare amplitude
$\bar a_{j}$, up to the constant $\mathrm{Ku}$. (The realised turnover time is
longer by $\sqrt{\ln2/\ln s}$ --- unity at $s=2$, $\sqrt2$ at $s=\sqrt2$; we tie
$\tau_{j}$ to $\bar a_{j}$ so that the time lattice is unchanged by the
refinement.)
The factor $\sqrt{\ln s/\ln2}$ makes the total energy and the spectrum independent
of $s$: refining from octaves ($s=2$) to half-octaves ($s=\sqrt2$) doubles the
number of bands and halves the energy in each, leaving
$\sum_{j}A_{j}^{2}$ and the second-order structure function unchanged to leading
order. At $s=2$ the normalisation is unity and one recovers the classical octave
model $\ell_{j}=\ell_{0}2^{-j}$, $A_{j}=2^{-j\alpha}$,
$\tau_{j}=\mathrm{Ku}\,\ell_{0}2^{-j(1-\alpha)}$. Why the refinement matters is a
measurement, reported in Section~\ref{sec:richardson}.

\paragraph{Profile.}   Set $b_j := \ell_j/2$. The only condition on the integral scale is $\ell_{0}\le L$, so that the
largest eddy fits in the torus; the runs use $\ell_{0}=L/2$. Fix then a specific ``model macroscopic eddy'', namely a non-trivial, zero-average, divergence-free velocity field
$\Wv: \mathbb R^n \to \mathbb R^n$ with the property that $\operatorname{supp}\Wv\subset[-\tfrac12,\tfrac12]^{n}$ and that its maximum modulus is normalised to $1$. We then define the model eddies at scale $j$ as
\begin{equation}\label{eq:profile}
\Wv^j (\xv) := \Wv (\ell_j^{-1} \xv)\, .
\end{equation}
With this normalisation
$\operatorname{supp}\Wv^{j}\subset[-b_{j},b_{j}]^{n}$ automatically. For instance in two space dimensions we could take
$w_{j}(\sigma)=\cos^{2}\!\bigl(\pi\sigma/2b_{j}\bigr)\mathbf 1_{\{|\sigma|<b_{j}\}}$, let $\bar \Wv^j:= \nabla^\perp w_j$, where $\nabla^{\perp}\!f=(\partial_{2}f,-\partial_{1}f)$, $v_j := \max |\bar{\Wv}^j|$, and set $\Wv^j := v_j^{-1} \bar{\Wv}^j$. In three dimensions we could use a vector potential $B_j$ in place of $w_j$, set $\bar{\Wv}^j = \nabla \times B_j$ and proceed analogously.
The following properties are used throughout the rest of the paper and nothing else:
$\operatorname{supp}\Wv^{j}\subset[-b_{j},b_{j}]^{n}$;
$\nabla\!\cdot\!\Wv^{j}\equiv0$; $\max |\Wv^j|=1$, $\Wv^{j}\in C^{0,1}$, and its average is $0$.

\paragraph{Epochs.} Generation $j$ is renewed on the time lattice $k\tau_{j}$,
$k\in\Zb$. Epoch $(j,k)$ lives on $|t-k\tau_{j}|<\tau_{j}$ with weight
\begin{equation}
g_{j,k}(t)=\cos^{2}\!\Bigl(\frac{\pi(t-k\tau_{j})}{2\tau_{j}}\Bigr)\,
\mathbf 1_{\{|t-k\tau_{j}|<\tau_{j}\}} .
\label{eq:window}
\end{equation}
Exactly two epochs are alive at any $t$ outside the discrete set
$\tau_{j}\Zb$, and the exponent $2$ is not cosmetic: writing
$\sigma=t/\tau_{j}-\lfloor t/\tau_{j}\rfloor$,
\begin{equation}
\sum_{k}g_{j,k}(t)=\cos^{2}\tfrac{\pi\sigma}{2}+\sin^{2}\tfrac{\pi\sigma}{2}=1 ,
\label{eq:pou}
\end{equation}
so the epochs form an exact partition of unity, which is what makes the identity of
Theorem~\ref{thm:band} uniform in $t$; a cross-fade $\cos^{p}$ with $p\neq2$ loses
Eq.~\eqref{eq:pou} and forces a compensating retuning of $\alpha$. Consecutive
epochs carry independent amplitudes and independent birth lattices, so the second
moment picks up $\sum_{k}g_{j,k}^{2}=1-\tfrac12\sin^{2}\pi\sigma$ instead: the
energy of generation $j$ retains a modulation of period $\tau_{j}$ between $1$ and
$\tfrac12$. This is a property of any renewal cross-fade with independent epochs
and is common to every field compared here. Write
$\mathcal K_{j}(t)=\{k:g_{j,k}(t)>0\}$.

\paragraph{Random coefficients.} Let $\sigma_{j}=\lfloor 0.7\,\ell_{j}\rceil$. The
birth positions of epoch $(j,k)$ are the  $n_{j}=m_{j}^{n}$ points,
$m_{j}=\lfloor N/\sigma_{j}\rceil$, of the periodic lattice
$\{(o_{1}+a\sigma_{j},\,o_{2}+b\sigma_{j})\bmod N\}$, with the offset drawn
uniformly on  $\{0,\dots,\sigma_{j}-1\}^{n}$ independently for each $(j,k)$; the
wrap keeps $n_{j}$ independent of the offset when $\sigma_{j}\nmid N$; we write them
 $\bar{\pv}^{(j,k)}_{i}$ and we call them birth lattice points. The amplitudes are independent across
$i$ and across $(j,k)$,
\begin{equation}
\eta^{(j,k)}_{i}=\gamma_{i}\,e^{\,\varsigma_{j}\xi_{i}-\varsigma_{j}^{2}},\quad
\gamma_{i},\xi_{i}\sim\mathcal N(0,1),\quad
\varsigma_{j}^{2}=c_{\mu}\,j\ln s ,
\label{eq:weights}
\end{equation}
so that $\Eb[\eta]=0$, $\Eb[\eta^{2}]=1$ and
$\Eb[\eta^{4}]=3\,s^{4c_{\mu}j}$: unit variance at every generation and a
flatness growing geometrically in $j$, which is the inter-generation intermittency
of a multiplicative cascade. Setting $c_{\mu}=0$ returns i.i.d.\ standard normal
amplitudes; nothing below depends on the choice beyond
$\Eb[\eta]=0$, $\Eb[\eta^{2}]=1$ and independence. That the \emph{arrangement} of
the centres is equally inessential is not an assumption but a measurement: replacing the randomly offset lattice of Eq.~\eqref{eq:birth} by a Poisson process of
the same density --- $g\equiv1$ against a Bragg peak at the spacing --- moves the
dispersion exponent by $0.0014$, sixty-six times less than switching off the transport
does (Fig.~\ref{fig:three}C).

\section{Co-motion: the inductive step}
\label{sec:comotion}

Write $\uv_{<j}=\sum_{m<j}\uv_{m}$ with $\uv_{<0}\equiv\mathbf 0$. The construction
is inductive in $j$: given every coarser generation, the centres of generation $j$
are  carried by the sum of the previous generations. More precisely, they solve the
system of ODEs
\begin{equation}
\frac{d}{dt}\,\pv^{(j,k)}_{i}(t) =\uv_{<j}\bigl(\pv^{(j,k)}_{i}(t),\,t\bigr),
\label{eq:centres}
\end{equation}
subject to the initial condition
\begin{equation}\label{eq:birth}
\pv^{(j,k)}_i ((k-1) \tau_j) = \bar{\pv}^{(j,k)}_i\, .
\end{equation}
In particular the centres start at the birth lattice points at the initial time $t=(k-1)\tau_{j}$. The centres need only be tracked in the time interval $[(k-1)\tau_{j},(k+1)\tau_{j}]$, the lifetime of epoch $(j,k)$. The ones of generation $0$
never move. Equivalently, and this is the form in which the construction is most
compactly stated, one may integrate Eq.~\eqref{eq:centres} as a flow. We can
consider the global flow $\Phi^{<j}_{s,t}$ of $\uv_{<j}$ which starts as the identity at time $(k-1) \tau_j$, and regard the solution of Eqs.~\eqref{eq:centres}--\eqref{eq:birth} as tracking the trajectories of a finite number of particles starting at the birth lattice points. More precisely we have
\begin{equation}\label{eq:centers-flow}
\pv^{(j,k)}_{i}(t)=\Phi^{<j}_{(k-1)\tau_{j},\,t}\bigl(\bar{\pv}^{(j,k)}_{i}\bigr) .
\end{equation}
This
continuous form is what we use, and it is what makes
Proposition~\ref{prop:residual} hold at every instant rather than only at epoch
boundaries.

The point of Eq.~\eqref{eq:centres} is Eq.~\eqref{eq:frame} without the
registers. A particle and an eddy of generation $j$ that are close enough to
interact are advected by the \emph{same} coarse field, so the coarse velocity
cancels from their relative motion up to the variation of $\uv_{<j}$ across the
eddy. Proposition~\ref{prop:residual} bounds that variation by a quantity of the order
of the eddy's own velocity: the sweep of the coarse field is removed, and what remains
is comparable to the eddy rather than to the sweep.

\section{Eulerisation: push the centres, not the field}
\label{sec:euler}

It remains to say how a transported set of centres generates a velocity. The naive
answer --- compose the pattern with the flow map --- is the one that fails, and the
reason it fails is the content of the construction.

Write the epoch's eddies as a convolution. With the atomic measure
$\mu=\sum_{i}\eta_{i}^{(j,k)}\delta_{\bar\pv_{i}^{(j,k)}}$, the random combination
$\sum_{i}\eta_{i}^{(j,k)}\Wv^{j}(\xv-\bar\pv^{(j,k)}_{i})$ is $\Wv^{j}*\mu$, and with a
potential $B_{j}$ such that $\Wv^{j}=\nabla\times B_{j}$ it is $\nabla\times(B_{j}*\mu)$.
To move it along the flow $\Phi$ of $\uv_{<j}$ one may advect the potential,
$(B_{j}*\mu)\circ\Phi^{-1}$, and take the curl. That is the route we do \emph{not}
take.

The key point of this paper is the other one: push the measure forward,
$\Phi_{\#}\mu=\sum_{i}\eta_{i}^{(j,k)}\delta_{\pv^{(j,k)}_{i}(t)}$, and re-convolve.
Since $\nabla\times(B_{j}*\Phi_{\#}\mu)=\Wv^{j}*\Phi_{\#}\mu$, the model eddy is moved
without being deformed, and in Fourier
$\widehat{\Wv^{j}*\Phi_{\#}\mu}=\widehat{\Wv^{j}}\cdot\widehat{\Phi_{\#}\mu}$ --- a
product, not a convolution. The two routes behave very differently.

\begin{lemma}[the dichotomy]\label{lem:dichotomy}
Let $\mu=\sum_{i}\eta_{i}\delta_{\pv_{i}}$ be the atomic measure of an epoch, let
$\Phi$ be a measure-preserving diffeomorphism of $\Tb^{n}$, and write
$\Psi=B_{j}*\mu$, so that $\Wv^{j}*\mu=\nabla\times\Psi$. The two objects
\[
\underbrace{\Psi\circ\Phi^{-1}}_{\text{compose the field}}
\qquad\text{and}\qquad
\underbrace{B_{j}*(\Phi_{\#}\mu)}_{\text{push the centres}}
\]
coincide for every $\mu$ whenever $\Phi$ is a translation, and differ as follows.
\begin{enumerate}
\item[(i)] \emph{Push the centres.} For every $\Phi$ and every $\mu$,
\begin{equation}
\begin{aligned}
\widehat{\Wv^{j}*(\Phi_{\#}\mu)}(\kv)&=\widehat{\Wv^{j}}(\kv)\,
\widehat{\Phi_{\#}\mu}(\kv),\\[2pt]
\bigl|\widehat{\Phi_{\#}\mu}(\kv)\bigr|&=\Bigl|\textstyle\sum_{i}\eta_{i}
e^{-i\kv\cdot\Phi(\pv_{i})}\Bigr| .
\end{aligned}
\label{eq:pushhat}
\end{equation}
The profile factor $\widehat{\Wv^{j}}$ is untouched by $\Phi$: the transform
factorises, and $\Phi$ enters the second factor only through unimodular phases. (It
is the factorisation that matters and not a support statement: $\Wv^{j}$ is
compactly supported, so $\widehat{\Wv^{j}}$ vanishes on no open set.)
\item[(ii)] \emph{Compose the field.} Let $\Theta\in H^{1}(\Tb^{n})$ be
arbitrary. Since $\det\nabla\Phi\equiv1$ one has
$\|\Theta\circ\Phi^{-1}\|_{L^{2}}=\|\Theta\|_{L^{2}}$, while
\begin{equation}
\bigl\|\nabla(\Theta\circ\Phi^{-1})\bigr\|_{L^{2}}^{2}
=\int_{\Tb^{n}}\bigl|(\nabla\Phi)^{-\top}\nabla\Theta\bigr|^{2},
\label{eq:h1}
\end{equation}
componentwise if $\Theta$ is vector-valued, and equality with
$\|\nabla\Theta\|_{L^{2}}$ holds for \emph{every} $\Theta\in H^{1}$ if and only if
$\nabla\Phi(\yv)\in O(n)$ for a.e.\ $\yv$.
\end{enumerate}
Unless $\nabla\Phi\in O(n)$ a.e., composition therefore moves the mean wavenumber
$\|\nabla\Theta\|_{L^{2}}/\|\Theta\|_{L^{2}}$ of some $\Theta$. Push-forward moves
it only through the structure factor, never through the profile --- and by
Theorem~\ref{thm:band} not even through the structure factor once the weights are
averaged over.
\end{lemma}

\begin{proof}
If $\Phi(\xv)=\xv+\cv$ then $B_{j}*(\Phi_{\#}\mu)(\xv)
=\sum_{i}\eta_{i}B_{j}(\xv-\pv_{i}-\cv)=\Psi\circ\Phi^{-1}(\xv)$. Part~(i) is
$\widehat{f*\mu}=\widehat f\,\widehat\mu$ with
$\Phi_{\#}\mu=\sum_{i}\eta_{i}\delta_{\Phi(\pv_{i})}$ and
$|e^{-i\kv\cdot\pv}|=1$. For~(ii), the chain rule gives
$\nabla(\Psi\circ\Phi^{-1})=(\nabla\Phi^{-1})^{\!\top}(\nabla\Psi)\circ\Phi^{-1}$;
substituting $\yv=\Phi^{-1}(\xv)$, whose Jacobian is $1$, gives
Eq.~\eqref{eq:h1}. A matrix $M$ with $\det M=1$ satisfies $|M^{-\top}\mathbf{v}|=|\mathbf{v}|$
for every $\mathbf{v}$ precisely when $M\in O(n)$. If $\nabla\Phi\notin O(n)$ on a
set of positive measure, pick a Lebesgue point $\yv_{0}$ of $\nabla\Phi$ there and a
direction $\ev$ with $|(\nabla\Phi(\yv_{0}))^{-\top}\ev|\neq|\ev|$; the oscillating
test functions $\Theta_{\varepsilon}(\yv)=\varepsilon\varrho(\yv)\cos(\ev\cdot
\yv/\varepsilon)$, with $\varrho$ a bump at $\yv_{0}$, concentrate
$\nabla\Theta_{\varepsilon}$ along $\ev$ as $\varepsilon\to0$ and make
Eq.~\eqref{eq:h1} differ from $\|\nabla\Theta_{\varepsilon}\|_{L^{2}}^{2}$.
\end{proof}

\begin{remark}
The same spreading can be written in Fourier variables: with
$\Phi^{-1}=\mathrm{id}-\mathbf D$,
\begin{equation}
\begin{aligned}
\widehat{\Psi\circ\Phi^{-1}}(\kv)&=\tfrac1{L^{n}}\textstyle\sum_{\qv}
\widehat\Psi(\qv)\,\mathcal T(\kv,\qv),\\[2pt]
\mathcal T&=\!\int\! e^{i(\qv-\kv)\cdot\xv-i\qv\cdot\mathbf D(\xv)}d\xv .
\end{aligned}
\label{eq:smear}
\end{equation}
If $\mathbf D$ is constant then
$\mathcal T(\kv,\qv)=L^{n}\delta_{\kv\qv}e^{-i\kv\cdot\mathbf D}$ and the modulus
is unchanged; if $\nabla\mathbf D\neq0$ the kernel is no longer diagonal. Turning
that into a width $O(|\qv|\,\|\nabla\mathbf D\|)$ needs hypotheses on
$\mathbf D$ --- $\mathbf D\in C^{1}$ with $\|\nabla\mathbf D\|_{\infty}<1$, so that
$\Phi$ is a diffeomorphism --- and a convention for what ``width'' means.
Part~(ii) needs neither, which is why the dichotomy is stated in a norm.
\end{remark}

\begin{remark}
Equality for every $\mu$ does not force $\Phi$ to be a translation: any
$\Phi=S+\cv$ with $S\in GL(2,\Zb)$ and $B_{j}\circ S=B_{j}$ --- for the $\cos^{2}$ example of
Section~\ref{sec:basis}, the eight symmetries of the square --- gives equality too,
being a measure-preserving automorphism of $\Tb^{2}$ with constant Jacobian. The
dichotomy the proof establishes is between $\nabla\Phi\in O(n)$
a.e.\ and not: a constant but non-orthogonal $\nabla\Phi$, such as a hyperbolic
automorphism of $\Tb^{2}$, already separates the two routes.
\end{remark}

\begin{remark}
The obvious repair of the composition route --- keep the translation and rotation
in $\nabla\Phi$ and delete the symmetric traceless part --- does not exist. A
matrix field $M=\mathsf I+\Omega$ with $\Omega$ antisymmetric and non-constant
violates the compatibility condition $\partial_{k}M_{ij}=\partial_{i}M_{kj}$ and is
therefore the Jacobian of no map: a ``locally rigid but spatially varying''
deformation is not available.
\end{remark}

Lemma~\ref{lem:dichotomy} says what to do. Transport the centres and re-convolve
with the same fixed profile.

\begin{definition}[the field]\label{def:main}
With the centres of Eq.~\eqref{eq:centres}, set
$\mu^{(j,k)}(t)=\sum_{i}\eta^{(j,k)}_{i}\delta_{\pv^{(j,k)}_{i}(t)}$,
and
\begin{equation}
\boxed{\begin{aligned}
\uv(\xv,t)=\sum_{j=0}^{J-1} A_{j}&\sum_{k\in\mathcal K_{j}(t)}g_{j,k}(t)\\
&\times\sum_{i}\eta^{(j,k)}_{i}\,\Wv^{j}\!\bigl(\xv-\pv^{(j,k)}_{i}(t)\bigr).
\end{aligned}}
\label{eq:field}
\end{equation}
\end{definition}

The velocity at $\xv$ at time $t$ is a finite sum of translated copies of $J$
fixed, compactly supported, divergence-free vector profiles, one term per live
eddy, with time-dependent centres and constant amplitudes. Because
$\operatorname{supp}\Wv^{j}\subset[-b_{j},b_{j}]^{n}$ only eddies with
$|\xv-\pv_{i}|_{\infty}<b_{j}$ contribute, so the sum is finite and local; in the
octave configuration ($s=2$, $J=6$, $N=512$) some $2\times10^{4}$ eddies are alive
in total, and $3\times10^{4}$ in the half-octave configuration used below.

Equations \eqref{eq:centres} and \eqref{eq:field} look circular --- the centres
move under a field built from the centres --- but the coupling is triangular.

Eqs.~\eqref{eq:centres}, \eqref{eq:birth} and \eqref{eq:field} have a unique solution
on $[0,\infty)$ whenever $B_{j}\in C^{1,1}$ for every $j$, constructed generation by
generation: at each step $\uv_{<j}$ is already built and Lipschitz, so the centres of
generation $j$ follow by Cauchy--Lipschitz and $\uv_{j}$ follows from them.

\begin{remark}[renewal is free]
Epoch $(j,k)$ is created at $t=(k-1)\tau_{j}$ on a fresh lattice and discarded at
$t=(k+1)\tau_{j}$. Both instants have $g_{j,k}=0$, so $\uv$ is continuous across
them: what is discarded was multiplying zero, and what is created multiplies zero.
\end{remark}

\section{Properties of the field}
\label{sec:props}

Two properties follow at once from the definition. For each $t$, $\uv(\cdot,t)$ is a
function on $\Tb^{n}$: every ingredient of Eq.~\eqref{eq:field} is fixed by $\xv$ and
$t$, the centres solve an ordinary differential equation in $t$ alone, and nothing
depends on the history of a particle passing through $\xv$, as it does in
Eq.~\eqref{eq:frame}. And $\nabla\!\cdot\!\uv\equiv0$ for every $t$ and every
configuration of the centres, since each generation is a curl.

\begin{proposition}[measure preservation; no folding]\label{prop:flow}
Let $\Phi_{s,t}$ be the flow of $\uv$. Then $\Phi_{s,t}$ is a bi-Lipschitz
homeomorphism of $\Tb^{n}$ preserving Lebesgue measure, with
$\det D\Phi_{s,t}=1$ a.e.; if $B_{j}\in C^{2}$ then $\Phi_{s,t}$ is a $C^{1}$
diffeomorphism and $\det D\Phi_{s,t}\equiv1$.
\end{proposition}
\begin{proof}
$\uv(\cdot,t)$ is Lipschitz, so the flow exists, is unique and is injective at every
$t$. Injectivity already says that no two fluid elements reach the same point, so the
flow cannot fold, and Liouville's formula with $\nabla\!\cdot\!\uv=0$ gives measure
preservation. The $C^{1}$ statement needs $B_{j}\in C^{2}$, which the $\cos^{2}$ example of
Section~\ref{sec:basis} is not ($w_{j}''$ jumps at $\pm b_{j}$); the bi-Lipschitz
statement suffices here.
\end{proof}

The per-particle construction behaves very differently. Of $384$ triads of particles
with side a quarter of the smallest eddy, $28\%$ fold under Eq.~\eqref{eq:frame},
against $0.01\%$ for the field of Definition~\ref{def:main} and $1.8\%$ for the
frozen field, which cannot fold either and so measures the interpolation floor.

\begin{theorem}[exact preservation of the expected spectral envelope]\label{thm:band}
For every $t$ and every configuration of the centres,
\begin{equation}
\widehat{\uv_{j}}(\kv,t)=A_{j}\,
\widehat{\Wv^{j}}(\kv)\sum_{k\in\mathcal K_{j}(t)}g_{j,k}(t)
\sum_{i}\eta^{(j,k)}_{i}e^{-i\kv\cdot\pv^{(j,k)}_{i}(t)},
\label{eq:bandhat}
\end{equation}
and, because the weights are i.i.d., zero-mean and independent of the centre positions
(generation $j$ is carried by $\uv_{<j}$, which is built from generations $m<j$
alone), the cross terms in the second moment vanish and
\begin{equation}
\Eb\bigl|\widehat{\uv_{j}}(\kv,t)\bigr|^{2}
=A_{j}^{2}n_{j}\bigl|\widehat{\Wv^{j}}(\kv)\bigr|^{2}
\!\!\!\sum_{k\in\mathcal K_{j}(t)}\!\!\!g_{j,k}(t)^{2}
\label{eq:bandexact}
\end{equation}
for every $\kv$, every $t$ and every trajectory of the centres. The $\kv$-dependence
is carried entirely by the fixed profile $\widehat{\Wv^{j}}$. The motion of the centres
enters only through the unimodular phases $e^{-i\kv\cdot\pv_{i}(t)}$, which change the
structure factor of the measure but not the spectral envelope, so the carrier cannot
redistribute generation $j$ across scales and no filter is needed. The expectation in
Eq.~\eqref{eq:bandexact} is over the generation-$j$ amplitudes at a fixed trajectory
of the centres; it is conditional on the coarse field rather than an average over it,
and neither equation says that the energy at a given $\kv$ is preserved in a single
realisation.
\end{theorem}
\begin{proof}
Eq.~\eqref{eq:bandhat} is $\widehat{\Wv^{j}*\mu}=\widehat{\Wv^{j}}\,\widehat\mu$.
For Eq.~\eqref{eq:bandexact}, expand the square: by
Eq.~\eqref{eq:weights} the weights satisfy
$\Eb[\eta^{(j,k)}_{i}\eta^{(j,k')}_{i'}]=\delta_{kk'}\delta_{ii'}$, and they are
independent of $\{\pv^{(j,k)}_{i}(t)\}$, so every off-diagonal term vanishes and
each of the $n_{j}$ diagonal terms contributes $|e^{-i\kv\cdot\pv_{i}}|^{2}=1$.
No hypothesis on the centres is used.
\end{proof}

Because $\Wv^{j}$ has compact support, its transform does not vanish on any open set,
so no generation is strictly confined to a band; what Eq.~\eqref{eq:bandexact} fixes
is the envelope. In particular the fraction of each generation's energy near its own
wavenumber is, on average, unchanged by the transport. Fig.~\ref{fig:one}C checks this
on the discretised field. With the centres transported as in Definition~\ref{def:main},
the fraction of each generation's energy inside $[0.7,1.4]\,k_{j}$, $k_{j}=2\pi/\ell_{j}$, averaged over
four realisations, agrees with that of the frozen field to within $0.022$, with no filter
applied. The composition route of Lemma~\ref{lem:dichotomy}, applied to the same
random data, leaves only $1.4$ to $7.3\%$ of each generation's energy in its band and lowers
the realised regularity from $0.369$ to $0.145$. Movie~S1 shows the two transports side by
side.

The centres also keep, on average, the density at which they were born. Since the flow
of $\uv_{<j}$ preserves area, the number of centres in a set $B$ equals the number of
birth lattice points in its preimage, and for any $B$ whose boundary has finite
$(n-1)$-dimensional upper Minkowski content this count is $|B|/\sigma_{j}^{n}$ up to a
boundary error, uniformly in $j$ wherever $\operatorname{Lip}(\Phi^{-1})$ is uniformly
bounded, as it is for bounded weights at small Kubo number, where the flow over one renewal
period and its inverse are Lipschitz with constant at most $2$ \cite{FeffermanMayboroda2026,delellis}. Individual centres can still approach one another, since an area-preserving
map compresses in one direction while it stretches in another, but this is harmless:
two coincident centres simply make one eddy, of amplitude $\eta_{1}+\eta_{2}$. The spectral envelope does not need even this much, because it is blind to the
arrangement: with every centre of an epoch at a single point,
$\Eb|\widehat{\mu^{(j,k)}}|^{2}=n_{j}$ still. The density of the centres matters only
for statistics beyond second order, including the Lagrangian ones of
Section~\ref{sec:richardson} (Appendix~\ref{app:centres}).

The per-particle construction satisfies the frame equation, Eq.~\eqref{eq:frame},
identically. The field of Definition~\ref{def:main} satisfies it only approximately,
and the error can be written down.

\begin{proposition}[residual sweeping]\label{prop:residual}
Let $\Xv(t)$ solve $\dot\Xv=\uv(\Xv,t)$ and let $\pv_{i}(t)$ be any centre of
generation $j$. Then
\begin{equation}
\begin{aligned}
\frac{d}{dt}\bigl[\Xv-\pv_{i}\bigr]
&=\underbrace{\uv_{\ge j}(\Xv,t)}_{\text{the exact frame, Eq.~\eqref{eq:frame}}}\\
&\quad+\underbrace{\bigl[\uv_{<j}(\Xv,t)-\uv_{<j}(\pv_{i},t)\bigr]}_{=:\ \mathbf e_{i}(t)},
\end{aligned}
\label{eq:residual}
\end{equation}
and, since $\Wv^{j}(\Xv-\pv_{i})=0$ unless $|\Xv-\pv_{i}|_{\infty}<b_{j}$, the
error is only ever evaluated where
$|\mathbf e_{i}|\le\sqrt2\,b_{j}\,\mathrm{Lip}(\uv_{<j})$. It vanishes identically
if $\uv_{<j}$ is a uniform translation.
\end{proposition}
\begin{proof}
Subtract Eq.~\eqref{eq:centres} from $\dot\Xv=\uv_{\ge j}(\Xv,t)+\uv_{<j}(\Xv,t)$; the
bound is the Lipschitz estimate on the support of $\Wv^{j}$.
\end{proof}

The residual changes the velocity against which generation $j$ decorrelates. Because
$A_{m}/\ell_{m}$ increases with $m$, the nearest coarser generation dominates the
Lipschitz constant, $\mathrm{Lip}(\uv_{<j})\le C_{s}A_{j-1}/\ell_{j-1}$ with
$C_{s}=(1-s^{-(1-\alpha)})^{-1}$. With $b_{j}=\ell_{j}/2$ and $A_{j-1}=s^{\alpha}A_{j}$
this gives, at $\alpha=1/3$,
\begin{equation}
|\mathbf e_{i}|\ \lesssim\ \tfrac{1}{\sqrt2}\,C_{s}\,s^{\alpha-1}A_{j}
=\begin{cases}1.2\,A_{j} & (s=2),\\[1pt] 2.7\,A_{j} & (s=\sqrt2).\end{cases}
\label{eq:errsize}
\end{equation}
The inequality holds for typical values and not deterministically, because the
supremum of $\nabla\uv_{<j}$ grows with the depth of the cascade through the
log-normal weights: at $N=512$ the r.m.s.\ of $\nabla\uv_{<j}$ stays at about twice
$C_{s}A_{j-1}/\ell_{j-1}$ for $j=4$, $7$ and $10$, while its supremum grows from $8$ to
$30$ times that value. With log-normal weights that growth is the obstacle to letting
$J\to\infty$ in the flow, although the field itself converges (Theorem~\ref{thm:L2-limit}); with bounded weights and a small Kubo
number it is absent, and the field is H\"older continuous uniformly in $J$ \cite{FeffermanMayboroda2026,delellis}.

The bound is of order $A_{j}$, so the residual is not small; what matters is what it
replaces. The sweeping removed by Eq.~\eqref{eq:centres} is dominated by the largest
generation, $(\sum_{m<j}A_{m}^{2})^{1/2}=O(A_{0})$, because $A_{m}$ decreases with $m$.
Co-motion therefore lowers the velocity that sets the decorrelation time of generation
$j$ from $O(A_{0})$ to $O(A_{j})$, a factor $s^{j\alpha}$, which is about three at the
smallest scales of our runs. If this exchange were complete, replacing the sweeping
time in Eq.~\eqref{eq:EB} by $r/A_{j}$ at $\ell_{j}\simeq r$ would give a pair
diffusivity $K\sim\varepsilon^{1/3}r^{4/3}$ in place of
$\varepsilon^{2/3}r^{5/3}/U_{0}$, and hence $\langle r^{2}\rangle\sim\varepsilon t^{3}$
in place of $t^{6}$. For bounded weights and a small Kubo number the conclusion is a theorem:
Ref.~\cite{FeffermanMayboroda2026} proves $c\,\mathrm{Ku}^{3}t^{3}\le\Eb R_{t}^{2}\le Ct^{3}$ at
$\alpha=1/3$ over a range of times, uniformly in $J$, and its central estimate is the one
above: the coarse flow carries the fine eddy together with the point at which it is read,
so only the increment of $\uv_{<j}$ across the eddy survives. For the log-normal weights
and order-one Kubo numbers used here the question is quantitative ---
Eq.~\eqref{eq:errsize} bounds the size of the residual, whereas the third entry of
Eq.~\eqref{eq:EB} is a correlation time, and a residual of amplitude $A_{j}$ could
still decorrelate on the much shorter time of the coarse field that produces it --- and
the pair measurements of the next section settle it.

The field itself, though not its Lipschitz constant, has a limit as generations are
added. The following result and its proof are due to L.~Sarieddine (private
communication).

\begin{theorem}[$L^{2}$ limit]\label{thm:L2-limit}
Extend the construction below the grid scale by giving every generation the birth
lattice of spacing $\sigma_{j}=0.7\,\ell_{j}$, with an offset drawn uniformly from
$[0,\sigma_{j})^{n}$. Then for every $t$ the fields $\uv^{J}=\sum_{j<J}\uv_{j}$ converge
in $L^{2}(\Tb^{n})$ almost surely as $J\to\infty$.
\end{theorem}

The idea is that each generation is small in $L^{2}$ on average. Generation $j$ is built
from the coarser ones alone, so the $\uv^{J}$ are partial sums of a single series. The
weights have zero mean and are independent of the centres, so the cross terms drop out
and $\Eb\|\uv_{j}\|_{L^{2}}\le CA_{j}$, with $C$ independent of $j$
because the centres sit $0.7\,\ell_{j}$ apart and $n_{j}\ell_{j}^{n}$ is bounded. Since
$A_{j}\propto s^{-j\alpha}$ is summable, $\sum_{j}\|\uv_{j}\|_{L^{2}}$ is finite almost
surely and the partial sums converge.

\begin{figure*}[t]
\centering
\includegraphics[width=\textwidth]{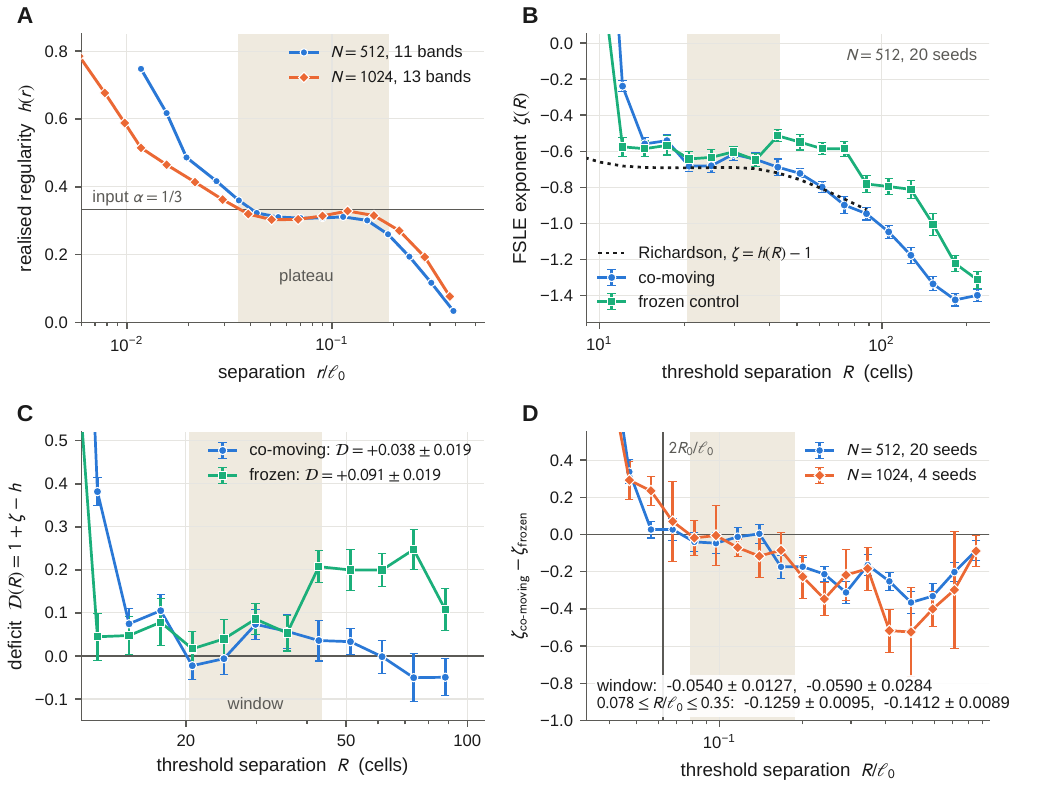}
\caption{\textbf{Richardson dispersion in the co-moving field.}
(\emph{A}) Realised regularity $h(r)=\tfrac12\,d\log S_{2}/d\log r$ at $N=512$ and
$N=1024$. Half-octave bands produce a plateau at $h\approx0.31$, near the input value
$\alpha=1/3$, which widens as the resolution increases.
(\emph{B}) The local exponent $\zeta(R)$ of the finite-size Lyapunov exponent for the
co-moving field and for the frozen control, both on the same random data, against the
Richardson line $\zeta=h(R)-1$ measured on the same field.
(\emph{C}) The deficit $\Dsw(R)=1+\zeta(R)-h(R)$. In the window (shaded)
the co-moving field gives $+0.038\pm0.019$, consistent with zero, and the frozen field gives
$+0.091\pm0.019$.
(\emph{D}) The paired difference between the two arms at both resolutions, against
$R/\ell_{0}$. Its change of sign near $2R_{0}/\ell_{0}$ is the release transient.}
\label{fig:two}
\end{figure*}

\section{Richardson dispersion}
\label{sec:richardson}

We now put particles in the field. Each one obeys
\begin{equation}
\dot\Xv=\uv(\Xv,t),
\label{eq:tracer}
\end{equation}
with $\uv$ the field of Definition~\ref{def:main}, and we follow pairs released a small
distance $R_{0}$ apart. The question is the one of Fig.~\ref{fig:one}A. In a real flow a
pair of size $R$ is pulled apart by the eddies of its own size and stays with them for
their turnover time. In a frozen synthetic field it is dragged through those eddies by the
large ones and loses them after the much shorter sweeping time $R/U_{0}$. We want to know
which of the two clocks the co-moving field keeps; Movie~S2 shows one pair in both fields. The computations are two-dimensional,
which is where the necessary range of scales is affordable (Appendix~\ref{app:methods}).

The dispersion measurements are read against a frozen control, the same field with
Eq.~\eqref{eq:centres} switched off, so that each eddy stays where it was born. The control is built from the
same random numbers as the experiment, down to the released pairs, and differs only in
whether the eddies move. The difference between the two is therefore the effect of
co-motion, and taking it realisation by realisation removes most of the scatter.

\subsection{Richardson's law at fixed scale}

The textbook form of the law, $\langle R^{2}\rangle\sim t^{3}$, is awkward to test over a
finite range of scales. At any given time some pairs are still close together while others
have already reached the largest eddies, and the average mixes the two. It is cleaner to
turn the question around, fix the separation and measure how long the pairs take to reach
it \cite{Artale1997,Aurell1997}. We set thresholds $R_{i}=\rho^{i}R_{0}$, record the first
time $T_{i}$ at which a pair exceeds each one, and form the finite-size Lyapunov exponent,
the rate at which pairs of size $R$ grow, together with its local slope,
\begin{equation}
\lambda(R_{i})=\frac{\ln\rho}{\langle T_{i}-T_{i-1}\rangle},\qquad
\zeta(R)=\frac{d\log\lambda}{d\log R} .
\label{eq:fsle}
\end{equation}
Richardson's argument then becomes a statement about $\zeta$. In a field of H\"older exponent $h$, a
pair at separation $R$ that is pulled apart by the eddies of its own size grows at the rate
$\delta u(R)/R\sim R^{h-1}$, so $\zeta=h-1$. At
$h=1/3$ this is $\lambda\sim R^{-2/3}$, the $t^{3}$ law read at fixed scale
\cite{BoffettaSokolov2002a}. The theorem of Ref.~\cite{FeffermanMayboroda2026}, recalled in
Section~\ref{sec:props}, is the same law read at fixed time: there $\Eb R_{t}^{1-\alpha}$
grows linearly in $t$.

We do not assume the value of $h$. We measure it on the same field, from the velocity
structure function $S_{2}$, as $h(R)=\tfrac12\,d\log S_{2}/d\log r$ at $r=R$, and compare
the two sides of $\zeta=h-1$ through the deficit
\begin{equation}
\Dsw(R)=1+\zeta(R)-h(R).
\label{eq:defect}
\end{equation}
Pairs that obey Richardson's law, at whatever regularity the field actually has, give
$\Dsw=0$. Sweeping makes $\Dsw$ positive: a pair dragged through a pattern that stands
still spends a time proportional to $R$ in each eddy instead of its turnover time, so its
separation rate falls off with $R$ more slowly than Richardson's law predicts.
Intermittency also shifts $\Dsw$, for an unrelated reason: the exponent $\zeta$ is built from
mean crossing times and $h$ from the mean square velocity increment, and the two weigh the rare
strong eddies differently. We measure this shift in the Discussion, by changing the
intermittency of the weights (Fig.~\ref{fig:four}C).

\subsection{Co-motion restores Richardson's clock}

The deficit means something only where $h$ is constant, and the classical octave-spaced
field has no such range: its realised $h(r)$ falls steadily with $r$ however many octaves
are added. Half-octave bands, which sample the same scales twice as finely with the same
spectrum
($s=\sqrt2$ in Eq.~\eqref{eq:scales}), produce a plateau at $h\approx0.31$, close to the
input $\alpha=1/3$, that widens with resolution (Fig.~\ref{fig:two}A). We use them for the
dispersion runs and read the deficit in a fixed window inside the plateau, above the
release transient (Appendix~\ref{app:details}). At $N=512$ and over twenty realisations,
\begin{equation}
\begin{aligned}
\Dsw^{\mathrm{co\text{-}moving}}&=+0.038\pm0.019,\\
\Dsw^{\mathrm{frozen}}&=+0.091\pm0.019 .
\end{aligned}
\label{eq:Don}
\end{equation}
The co-moving deficit is consistent with zero and the frozen one is not
(Fig.~\ref{fig:two}C). The two arms share every amplitude, so the difference
\begin{equation}
\Dsw^{\mathrm{frozen}}-\Dsw^{\mathrm{co\text{-}moving}}=+0.052\pm0.014\quad(3.8\sigma)
\label{eq:excess}
\end{equation}
isolates what co-motion removes. With weaker intermittency, close to that of Navier--Stokes
turbulence, it doubles (Fig.~\ref{fig:four}C). Fig.~\ref{fig:two}B shows the same thing in the exponents themselves: the
co-moving field follows the Richardson line $\zeta=h(R)-1$ threshold by threshold, and the
frozen field rises above it.

\begin{figure*}[t]
\centering
\includegraphics[width=\textwidth]{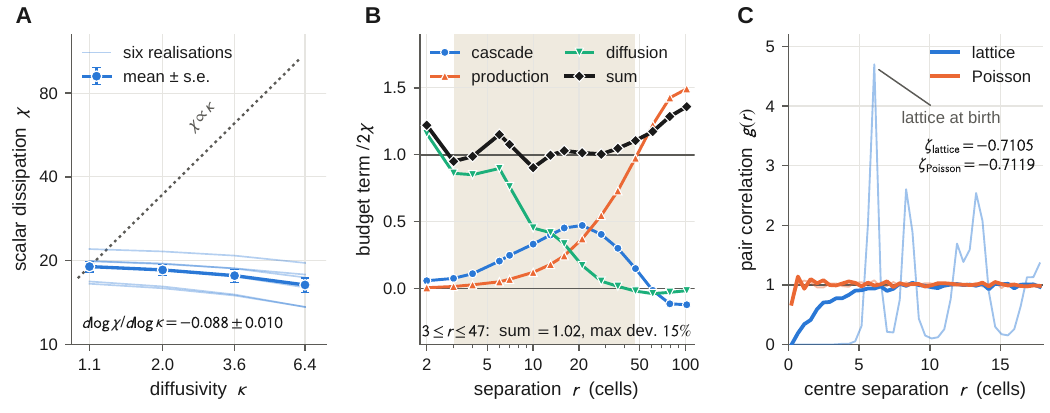}
\caption{\textbf{The scalar, and the eddy arrangement.}
(\emph{A}) Scalar dissipation $\chi=2\kappa\langle|\nabla\phi|^{2}\rangle$ against
diffusivity at fixed field, for six realisations of the co-moving field (thin lines)
and their mean (symbols, with standard errors). The dotted line, $\chi\propto\kappa$, is the behaviour of a scalar without an anomaly.
(\emph{B}) The scale-by-scale budget, Eq.~\eqref{eq:budget}, in the octave-spaced field:
three of its four terms, each measured independently and normalised by $2\chi$, and
the sum of all four (the unsteady term is below $1\%$ and is not drawn). The rise above
$r\approx50$ is the large-scale damping, which is not plotted.
(\emph{C}) Pair correlation of the eddy centres for a randomly offset lattice and for
a Poisson process of the same density. The dispersion exponents of the two differ by
$0.0014$.}
\label{fig:three}
\end{figure*}

Doubling the resolution does not change the result. At $N=1024$, on four independent
realisations, every number agrees with its $N=512$ value to within one standard deviation
(Fig.~\ref{fig:two}D; Table~\ref{tab:res}), although four realisations confirm the effect rather than
establish it on their own. Nor does the lifetime of the eddies. At Kubo numbers from
$0.25$ to $2$ the co-moving deficit stays consistent with zero, while the
frozen field, once the release transient has passed, separates pairs more slowly the longer
its eddies live, at $0.96$ of the co-moving rate at $\mathrm{Ku}=0.25$ and $0.51$ at
$\mathrm{Ku}=2$ (Table~\ref{tab:kubo}). That trend is the obstruction at work: the longer a frozen eddy
lives, the more likely a pair is to be swept out of it before it dies. The
regular arrangement of the centres plays no part either. By Theorem~\ref{thm:band} it is
invisible to second-order statistics, and in octave-spaced runs replacing the lattice by a
Poisson process of the same density moves the dispersion exponent by $0.0014$, sixty-six
times less than switching off the co-motion there (Fig.~\ref{fig:three}C). Richardson scaling in this field comes from the
motion of the eddies and not from where they sit.

\subsection{Spontaneous stochasticity}
\label{sec:spont}

Richardson's law has a sharper relative. If $\langle R^{2}\rangle\sim t^{3}$ holds with
no memory of $R_{0}$, then two particles that start from the same point still separate,
and their trajectories are not unique. This is spontaneous stochasticity, the property that
Drivas and Eyink showed to be equivalent to anomalous dissipation \cite{DrivasEyink2017}. The two
are not the same thing. Spontaneous stochasticity fixes no exponent: it can hold in a field
of any H\"older exponent $h<1$, with $\langle R^{2}\rangle\sim t^{2/(1-h)}$. And a power law
does not establish it, since over a finite range of scales a smooth field can imitate one.
We therefore test it directly.

We release the two members of a pair from the same point and give each its own weak
Brownian kicks $\boldsymbol{\beta}$,
\begin{equation}
d\Xv=\uv(\Xv,t)\,dt+\sqrt{2\kappa_{p}}\;d\boldsymbol{\beta} ,
\label{eq:spont}
\end{equation}
and record their separation $R(T)$ at a fixed time $T$. In a smooth flow the separation is
the noise, stretched by the flow, and the whole distribution of $R(T)$ shrinks like
$\kappa_{p}^{1/2}$ as the noise is turned down. If trajectories are not unique, the
distribution of $R(T)$ tends instead to a non-trivial limit as $\kappa_{p}\to0$
\cite{BGK1998}. Our field is Lipschitz below its smallest eddy, of size $\ell_{\min}$, so at finite $J$ its flow is unique
and $R(T)$ must eventually shrink; what we measure is how weakly it depends on $\kappa_{p}$,
and whether that dependence goes away as $\ell_{\min}$ is lowered. The noise acts on the
particles and not on a field, so there is no small scale to resolve and $\kappa_{p}$ can be
varied over many decades (Movie~S3).

In the smooth control, a single generation of eddies, the median separation scales as
$\kappa_{p}^{1/2}$ to four digits over twenty-two decades, from $\kappa_{p}=10^{-24}$ to
$10^{-2}$. With the eleven generations of the $N=512$ field it scales as
$\kappa_{p}^{0.0333\pm0.0038}$: over the same twenty-two decades it moves only from $40$ to
$279$ cells, where in the smooth control it changes by a factor $10^{11}$. The dependence that is left has a simple
origin. Below $\ell_{\min}$ a pair grows exponentially from the size set by the noise, so
the time it needs to escape into the rough range grows only as the logarithm of
$1/\kappa_{p}$, and that time shrinks as generations are added. So does the exponent, as
$\ell_{\min}^{0.47\pm0.05}$ (Table~\ref{tab:ladder}), and it vanishes as the cutoff is removed. Over the
upper five decades, $\kappa_{p}=10^{-7}$ to $10^{-2}$, where the pairs have reached the
integral scale by time $T$, the exponent is $+0.0128\pm0.0010$ and, across the three
deepest cascades, falls as $\ell_{\min}^{0.63\pm0.10}$, the escape-time prediction $\ell_{\min}^{1-\alpha}$.

The equivalence with anomalous dissipation is a statement about trajectories run backward
in time: the scalar at time $t$ is an average of its initial values over the paths that
arrive at each point at $t$. The two directions need not behave alike, as Burgers
turbulence shows \cite{EyinkDrivas2015}, so we replayed the stored field backward over the
five-decade window. The exponent is $+0.0132\pm0.0010$ backward and $+0.0128\pm0.0010$
forward, and the result holds in the direction the equivalence requires. Pairs run backward are also the route of the proof
of anomalous dissipation in Ref.~\cite{FeffermanMayboroda2026}, which bounds the separation of diffusing pairs
in the time-reversed field from below at Richardson's rate.

\section{Yaglom's relation and the dissipative anomaly}
\label{sec:anomaly}

Because $\uv$ is a velocity field and not a rule for moving particles, a passive scalar
can be placed in it. We solve
\begin{equation}
\partial_{t}\phi+\uv\!\cdot\!\nabla\phi=\kappa\Delta\phi+G\,u_{1},
\qquad
\chi=2\kappa\bigl\langle|\nabla\phi|^{2}\bigr\rangle ,
\label{eq:scalar}
\end{equation}
in the field of Definition~\ref{def:main}, pseudospectrally (Appendix~\ref{app:methods}). The source
$G\,u_{1}$ comes from an imposed mean gradient $G$ and feeds scalar variance in, mostly at the
largest scales; $\chi$ is the rate at which diffusion destroys it at the smallest, and in
between the flow has to carry it down through every scale. We ask two things of the
scalar: that this transfer obey Yaglom's law, and that the rate of destruction survive as
$\kappa\to0$.

\subsection{Yaglom's relation}

Yaglom's law \cite{Yaglom1949} says that the flux of scalar variance through scale $r$
equals the rate at which variance is destroyed. The flux is measured by the mixed
third-order structure function $S_{L}(r)=\langle\delta u_{L}(\delta\phi)^{2}\rangle$, and the
exact scale-by-scale budget in two dimensions reads
\begin{align}
&-\frac1r\partial_{r}\bigl[r\,S_{L}\bigr]
+2G\langle\delta u_{1}\delta\phi\rangle
+2\kappa\nabla^{2}_{r}\langle(\delta\phi)^{2}\rangle
-\partial_{t}\langle(\delta\phi)^{2}\rangle\notag\\
&\hspace{3.2cm}=2\chi+\mathcal S(r),
\label{eq:budget}
\end{align}
where $\nabla^{2}_{r}=r^{-1}\partial_{r}(r\partial_{r})$ and $\mathcal S(r)$ is a
damping of the two lowest wavenumber shells, which keeps the run statistically steady and
acts only at scales comparable to $\ell_{0}$.

Each term on the left can be measured on its own, and we measured all four, from $300$
snapshots of a stationary run of the octave-spaced field. Over $1.2$ decades of separation
their sum equals the independently measured $2\chi$ to within $15\%$, with no adjustable
constant (Fig.~\ref{fig:three}B). The cascade term carries the variance through the middle
of this range, the mean gradient takes over at large separations and diffusion at small
ones. Neither numerical diffusion nor a lack of stationarity can produce this balance:
$S_{L}$ is a signed third-order moment, the unsteady term is measured separately, and
numerical diffusion would show up as a failure to close. The balance also needs velocity and
scalar increments taken at the same point and the same instant, which only a velocity
field can supply.

In the inertial--convective range the budget reduces to Yaglom's law in its textbook form,
$S_{L}(r)=-\chi r$ \cite{Yaglom1949,Falkovich2001}, which has been observed in direct
numerical simulations of the Navier--Stokes equations
\cite{GotohWatanabe2015,Saito2024,IyerYeung2014}. A mean gradient injects variance at every
scale, so to see this form we replace it by a random source acting only on the lowest
wavenumber shell, and run at $N=4096$. Even so the measured flux falls short of $\chi r$,
by $18\%$ at the smaller diffusivity and $30\%$ at the larger, because at this resolution diffusion
still carries part of the variance through the range. That diffusive transfer is measured
too, so it can be put back, with two small corrections that do not depend on $\kappa$
(Appendix~\ref{app:details}). Pooling four realisations and two diffusivities, we then find
\begin{equation}
\Bigl\langle\frac{-S_{L}(r)}{r\chi}\Bigr\rangle_{\rm corrected}=0.9995\pm0.0018 ,
\label{eq:plateauval}
\end{equation}
within $0.3\sigma$ of the value Yaglom's law requires, and the dependence on $\kappa$ that
remains is more than forty times smaller than before the correction.

The scalar's own scaling exponent is a harder test. Obukhov and Corrsin tie it to the
velocity's through $h+2\beta=1$, where $\langle(\delta\phi)^{2}\rangle\sim r^{2\beta}$. At
our resolution the scalar's scaling range is squeezed between the diffusive and the forcing
scales, and $\beta$ has not yet reached the Obukhov--Corrsin value, but comparing two
resolutions shows it moving towards that value as the range widens (Appendix~\ref{app:details}).

\subsection{The dissipation does not follow the diffusivity}

The anomaly concerns the limit $\kappa\to0$. As the diffusivity is lowered, diffusion acts
at ever finer scales, but the flow keeps delivering variance to them at a rate it sets
itself, so $\chi$ need not follow $\kappa$ down. We probe this with a scan in $\kappa$ at
fixed field under the mean gradient of Eq.~\eqref{eq:scalar} (Fig.~\ref{fig:three}A and
Movie~S4). The
dissipation rate barely responds,
\begin{equation}
\frac{d\log\chi}{d\log\kappa}=-0.088\pm0.010 ,
\label{eq:anomaly}
\end{equation}
where a scalar without an anomaly would give $+1$. Over a factor $5.8$ in $\kappa$, $\chi$
changes by $14\%$, and the points are fitted much better by $\chi=\chi_{0}-B\kappa$, which
has a finite limit, than by a power law; $\chi_{0}$ is finite in every realisation. For a
version of this field with bounded symmetric weights and a small Kubo number the anomaly is
a theorem \cite{FeffermanMayboroda2026,delellis}: a decaying scalar loses, on average, at least a fixed fraction of its variance in a
fixed time as the diffusivity goes to zero, both when the diffusivity is lowered together
with the cutoff and in the limiting field itself.

The scan has to use the mean gradient. With the random source of the previous
subsection the injection rate is fixed in advance, so in a steady state $\chi$ equals it
whatever $\kappa$ is, and such runs say nothing about the limit. Under a mean gradient the
injection rate is an output of the flow, and only then does a $\kappa$-independent $\chi$
say something about the transport (Appendix~\ref{app:details}).

Co-motion matters for the scalar too. In paired runs of the octave-spaced field the scalar
variances of the frozen and co-moving fields differ by $0\pm4\%$, but the co-moving field
destroys variance faster: $\chi/\langle\phi^{2}\rangle$ is larger in each of four realisations,
by $20\pm4\%$ on average. The co-motion that restores Richardson's clock for pairs also
speeds up the scalar cascade, as one expects if both are set by how long a fluid element
stays with its own eddies, and the construction is what allows the two to be measured in
one object.

The scalar is finally destroyed at the Batchelor scale $\ell_{B}=\sqrt{\kappa/s_{\rm rms}}$,
with $s_{\rm rms}$ the r.m.s.\ strain rate. That scale lies below the smallest eddy,
$\ell_{\min}$, where the velocity is smooth. This does not undermine the measurement. What has to survive $\kappa\to0$ is the flux through
the scaling range, and Fig.~\ref{fig:three}B places it above $\ell_{\min}$: the cascade term
peaks near $r\approx21$ cells against $\ell_{\min}=8$, and only the diffusive term, which
dominates below $r\approx13$, depends on $\ell_{B}$. A rough velocity above a cutoff with a
Batchelor range below it is also the arrangement of Navier--Stokes turbulence at large
Schmidt number, the regime covered by the rigorous results of
Refs.~\cite{BBPS_JEMS2022,BBPS_CPAM2022}. The Lagrangian test of spontaneous stochasticity
has no Batchelor scale at all.

\section{Discussion}
\label{sec:discussion}

\subsection{What has been constructed}

Equation~\eqref{eq:field} defines, for any $\alpha\in(0,1)$, any number of generations
and any scale ratio, a divergence-free velocity field on $\Tb^{2}$ that is an honest
function of position, whose flow is measure- and orientation-preserving, whose
transport leaves the expected spectral envelope of each generation unchanged
(Theorem~\ref{thm:band}), and in which the
quasi-Lagrangian cancellation holds up to the error of Eq.~\eqref{eq:errsize}. In it,
pair separation obeys Richardson's law in the sharp local form $\zeta(R)=h(R)-1$,
threshold by threshold within the scaling range, at two resolutions and at Kubo numbers
from $0.25$ to $2$, with a residual consistent with zero, also at the intermittency of
Navier--Stokes turbulence, while the same field with its eddies held still carries an additional
deficit, $+0.052\pm0.014$, which doubles to $+0.102\pm0.016$ at that intermittency; for
bounded weights and a small Kubo number, Richardson's law is a theorem for a version of this
field, and so, for symmetric weights, is anomalous dissipation \cite{FeffermanMayboroda2026,delellis}. A passive scalar
in the field of Definition~\ref{def:main}
satisfies Yaglom's scale-by-scale relation, all four terms measured independently and
balancing to $15\%$ over $1.2$ decades, and reaches the inertial--convective
asymptote $S_{L}=-\chi r$ to $0.9995\pm0.0018$ once the measured diffusive transfer is
restored --- the coefficient fixed to $0.2\%$ with no adjustable constant, though
$h+2\beta=1$ is not reached in this range and we measure its approach rather than claim
it. Its dissipation rate barely follows the diffusivity, a test that needs an
advection--diffusion equation and is therefore unavailable to any per-particle
construction.

Sharper still, the field approaches spontaneous stochasticity as its cutoff is removed.
At any finite $J$ it is Lipschitz below $\ell_{\min}$ and its trajectories are unique,
but pairs released from a single point with independent Brownian kicks depend on the
particle diffusivity only through the time needed to cross that smooth range:
$d\log R_{\rm med}/d\log\kappa_{p}=+0.0333\pm0.0038$ over twenty-two decades, against
$1/2$ for a single-scale smooth control on the same random data, and this residual
falls as $\ell_{\min}^{0.47\pm0.05}$ as generations are added. Over the five decades in
which the pairs reach the integral scale the exponent is $+0.0128\pm0.0010$, falling
across the three deepest generation counts as $\ell_{\min}^{0.63\pm0.10}$, the
escape-time prediction $\ell_{\min}^{1-\alpha}$; run backward in time, the direction in
which Ref.~\cite{DrivasEyink2017} states the equivalence, it returns
$+0.0132\pm0.0010$. This is the sharper of the two Lagrangian statements: a field that
is smooth on the scales a pair explores can imitate a power law over a bounded window,
but its separation keeps the memory of $\kappa_{p}$, and it is non-uniqueness rather
than the exponent for which the equivalence is stated. We
are not aware of another synthetic velocity field with a finite correlation time --- as
opposed to a per-particle prescription, or a white-in-time ensemble in which sweeping
is absent by construction \cite{Kraichnan1968,BGK1998} --- in which the
Thomson--Devenish obstruction has been shown absent by direct measurement against a
control sharing its random data.

The mechanism is a single algebraic fact. A spectrally localised pattern $B_{j}*\mu$
can be moved in two inequivalent ways, and only one commutes with the band structure
(Lemma~\ref{lem:dichotomy}). Composing the field with a flow map is what a Lagrangian
intuition suggests, and it destroys the cascade; pushing the centres forward keeps it.
The distinction is invisible at the level of a translation, which is presumably why it
has not been isolated before: the two agree exactly for any translation, however
large, and it is the \emph{gradient} of the displacement --- the strain of the carrier,
the thing one is trying to model --- that separates them.

\subsection{Intermittency}

\begin{figure*}[t]
\centering
\includegraphics[width=\textwidth]{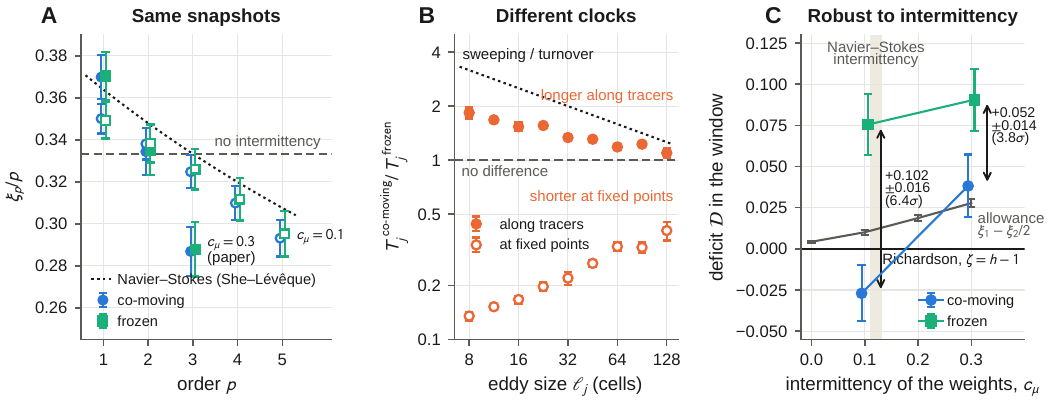}
\caption{\textbf{Same snapshots, different clocks.}
(\emph{A}) Exponents of the velocity structure functions,
$\langle|\uv(\xv+\mathbf{r},t)-\uv(\xv,t)|^{p}\rangle\propto r^{\xi_{p}}$, divided by $p$ and fitted
over the scaling range of Fig.~\ref{fig:two}A, for the co-moving field and the frozen control
with $c_{\mu}=0.3$ (filled, as everywhere else in this paper) and $c_{\mu}=0.1$ (open) in
Eq.~\eqref{eq:weights}; $N=512$, eight realisations each, converged orders only. Dotted: the
She--L\'ev\^eque exponents of Navier--Stokes turbulence \cite{SheLeveque1994}.
(\emph{B}) Decorrelation time of the velocity of generation $j$ in the co-moving field divided
by that in the frozen control, followed along tracers (filled) and at fixed points (open),
paired realisation by realisation ($N=512$, $c_{\mu}=0.3$, four realisations). Dotted:
$(\ell_{0}/\ell)^{1/3}$, the ratio of the sweeping time to the turnover time.
(\emph{C}) The deficit $\Dsw$ of Eq.~\eqref{eq:defect} in the window of Section~\ref{sec:richardson}
against $c_{\mu}$, for twenty seed pairs at each value, with the paired excess of the frozen field
and the Eulerian estimate $\xi_{1}-\xi_{2}/2$ over the same window. The band marks the
$c_{\mu}$ at which the exponents of \emph{A} match She--L\'ev\^eque. Details in Appendix~\ref{app:details}.}
\label{fig:four}
\end{figure*}

The weights of Eq.~\eqref{eq:weights} make the field intermittent, and its structure functions
measure how much (Fig.~\ref{fig:four}A). With $c_{\mu}=0.3$, $\xi_{p}/p$ falls from
$0.370\pm0.011$ at $p=1$ to $0.287\pm0.012$ at $p=3$, a drop of $0.083\pm0.003$, nearly three
times that of the She--L\'ev\^eque exponents of Navier--Stokes turbulence, $0.031$
\cite{SheLeveque1994}. This is a choice and not a prediction of the model: the drop is $0.010$,
$0.025$, $0.051$ and $0.083$ for $c_{\mu}=0$, $0.1$, $0.2$ and $0.3$, and $c_{\mu}\simeq0.12$
reproduces Navier--Stokes. All of them pass through $\xi_{2}/2\simeq\alpha=1/3$ at $p=2$,
because the spectrum is prescribed. Navier--Stokes turbulence is pinned instead at $p=3$ by the
four-fifths law; here nothing holds $\xi_{3}$ at one, since nothing carries energy between
scales, and it falls from $1.007\pm0.022$ with Gaussian weights to $0.861\pm0.035$ at $c_{\mu}=0.3$.
In a snapshot the co-moving field and the frozen control agree to $0.002$ at every order, as
they must, since they share every amplitude and co-motion only moves the eddies. Where they
differ is in time (Fig.~\ref{fig:four}B). Along a tracer the velocity of one generation stays
correlated $1.1$ to $1.8$ times longer in the co-moving field, the more so the smaller the
eddies, while at a fixed point it decorrelates $2.5$ to $7$ times faster, because the eddies are
now carried past it. The tracer gain falls short of the ratio $(\ell_{0}/\ell)^{1/3}$ of the
sweeping to the turnover time, since a tracer moves relative to its eddy with the velocity of
that generation and the finer ones and with the strain of the coarser ones
(Proposition~\ref{prop:residual}).

Intermittency also enters the Richardson test of Section~\ref{sec:richardson}, because $\zeta$ is
built from mean exit times and $h$ from $S_{2}$. To see how much, we repeated the paired
experiment with $c_{\mu}=0.1$, close to Navier--Stokes, on the same twenty seeds
(Fig.~\ref{fig:four}C). The frozen field still lags, and by more: the excess
$\Dsw^{\mathrm{frozen}}-\Dsw^{\mathrm{co\text{-}moving}}$ is $+0.102\pm0.016$ ($6.4\sigma$),
against $+0.052\pm0.014$ ($3.8\sigma$) at $c_{\mu}=0.3$. The co-moving deficit is at most two
standard errors from zero at both values, $-0.027\pm0.017$ and $+0.038\pm0.019$. How the deficits
move with $c_{\mu}$ is less simple. The frozen one moves as the Eulerian estimate
$\Dsw\simeq\xi_{1}-\xi_{2}/2$ predicts, by $+0.015\pm0.015$ against $+0.019$; the co-moving one
moves by $+0.065\pm0.015$, more than three times the $+0.018$ of that estimate. We think this is
because pairs that move with the eddies stay with the strong ones, so the rare strong eddies
weigh more on their exit times than the statistics of a snapshot allow. The residual deficit of
the co-moving field is therefore not the intermittency allowance of a monofractal argument.
What is robust is that it is consistent with zero, with the intermittency of the model and with
that of Navier--Stokes turbulence, while the same field with its eddies held still is not.

\subsection{No cascade, and the origin of the arrow of time}
\label{sec:nocascade}

The field has no energy cascade, in either direction. Its velocity is prescribed, the energy
of each generation is fixed by construction, and no equation moves energy between scales.
The usual diagnostics of a cascade agree. In both arms the skewness of the longitudinal
velocity increment, $\langle\delta u_{L}^{3}\rangle/\langle\delta u_{L}^{2}\rangle^{3/2}$, is
below $0.03$ in magnitude at every separation from $2$ to $64$ cells, where the four-fifths
law gives about $-0.3$ in the inertial range of three-dimensional turbulence. The filtered
energy flux $\Pi_{\ell}$ averages to less than $5\%$ of its mean magnitude at $\ell=16$, $32$
and $64$ cells, and it is negative, carrying energy to larger scales, on $49$ to $53\%$ of the
area (Appendix~\ref{app:details}). Energy is moved up and down the scales in equal measure and nowhere on
balance. In the frozen field this is exact, because reversing the sign of every weight leaves
the ensemble unchanged and reverses $\uv$; co-motion breaks that symmetry without producing a
measurable cascade. The scalar, by contrast, cascades forward, as Yaglom's law requires.
Richardson dispersion and the dissipative anomaly in this field do not come from a cascade of
energy but from the roughness of the field and the co-motion of its eddies.

Nor is the velocity measurably irreversible. The frozen ensemble is invariant under
$t\to-t$, $\uv\to-\uv$, since the cross-fade and the weights are symmetric. The co-moving
ensemble is not exactly invariant: every epoch is born on a lattice and carried forward, so
its arrangement melts in one direction of time. But the arrangement barely enters the
dispersion (Section~\ref{sec:richardson}), and pairs separate alike in the two directions,
with spontaneous-stochasticity exponents $+0.0128\pm0.0010$ forward and $+0.0132\pm0.0010$
backward. In three-dimensional Navier--Stokes turbulence the cascade gives the velocity an
arrow of its own: pairs separate faster backward than forward \cite{Berg2006}, and fluid
particles lose kinetic energy faster than they gain it \cite{Xu2014}. This field has neither.

The arrow of time of the anomaly comes from the scalar. For any $\kappa>0$ the scalar problem
is irreversible, but in a velocity field whose trajectories are unique the variance lost in a
fixed time vanishes with $\kappa$. By the Lagrangian fluctuation--dissipation relation
\cite{DrivasEyink2017}, the scalar at $(\xv,t)$ is the average of its initial values over the
starting points of noisy trajectories run backward from $(\xv,t)$, and the variance it has lost
is set by how widely those starting points are spread. Spontaneous stochasticity keeps them
spread over a finite region as $\kappa\to0$, so the scalar stays an average of its initial
values over a finite patch and loses variance at a finite rate. Time-reversal symmetry is
therefore broken spontaneously. At $\kappa=0$ the equations are reversible and the velocity
nearly so, yet the limit of the solutions at $\kappa>0$ picks a direction of time, as a
symmetry is picked by a field that has been switched off. Reversing the velocity does not
unmix the scalar, because the reversed problem, at any $\kappa>0$, mixes forward in its own
time. The Kraichnan model, whose white-in-time Gaussian velocity is exactly symmetric under
time reversal, has the same anomaly \cite{BGK1998,Falkovich2001}; here the same mechanism
operates in a velocity with finite correlation times and the sweeping structure of a physical
flow. The arrow of time of turbulent mixing does not need a velocity that cascades. It needs
trajectories that are not unique.

\subsection{Relation to the rigorous constructions}

The mathematical literature has lately produced velocity fields for which anomalous
scalar dissipation is a theorem. Colombo, Crippa and Sorella
\cite{ColomboCrippaSorella2023} obtain the full supercritical Obukhov--Corrsin range
with a uniform-in-$\kappa$ H\"older bound, and show that vanishing diffusivity fails to
select a solution; on the other side of the threshold $\phi\in C^{\beta}$ with
$\beta>(1-\alpha)/2$ forbids the anomaly outright \cite{DEIJ2022}. Armstrong and Vicol
\cite{ArmstrongVicol2025} construct an explicit deterministic field in
$C^{0}_{t}C^{0,\alpha}_{x}$, $\alpha<1/3$, for which every $H^{1}$ datum dissipates, by
a fractal-homogenisation cascade inserting periodic shears at infinitely many scales
\emph{in Lagrangian coordinates}; Eyink \cite{Eyink2024} identifies it as the first
synthetic velocity field carrying the sweeping structure of a physical flow. Burczak,
Sz\'ekelyhidi and Wu \cite{Burczak2023} combine that machinery with convex integration
to obtain the same for scalars advected by $C^{1/3-}$ weak solutions of Euler.

The field of Definition~\ref{def:main} is complementary. It converges in $L^{2}$ as
$J\to\infty$ (Theorem~\ref{thm:L2-limit}), and \cite{FeffermanMayboroda2026,delellis} prove Richardson's law for its flow uniformly in $J$, and
anomalous dissipation for a passive scalar it advects, the latter both for the truncated
fields and for their limit. They do so for a version with abrupt renewals in place of the
cross-fade, bounded weights (symmetric, for the dissipation), a radial $C^{4}$ stream function for the eddy and a
small Kubo number. The proof needs one property of the profile, its
condition (ND): for a single eddy, the variance of the velocity increment across the
separation, $D_{N}(r)$, is at least the one along it, $D_{L}(r)$. The bump used in our
runs, $\cos^{2}(\pi x_{1}/2b_{j})\cos^{2}(\pi x_{2}/2b_{j})$, is neither radial nor $C^{4}$,
so we checked this direction by direction: $D_{N}\ge D_{L}$ holds in every direction to
grid accuracy, with $D_{N}-D_{L}\propto r^{2}$ near $r=0$ (Appendix~\ref{app:methods}). The
measurements here extend the picture to the log-normal weights and the cross-fade of
Definition~\ref{def:main}, with pair dispersion measured at Kubo numbers from $0.25$ to $2$.
The field is also explicit, cheap, tunable in $\alpha$, and its Lagrangian statistics can be read off
directly. It realises the sweeping
structure by a different device: where Armstrong and Vicol insert each scale in the
Lagrangian coordinates of the preceding ones, Eq.~\eqref{eq:centres} advects the
\emph{generators} and re-convolves, which is what makes Theorem~\ref{thm:band} exact
rather than asymptotic.
The obstruction to $J=\infty$ is visible in Eq.~\eqref{eq:errsize}, a residual of order
$A_{j}$ per generation whose ratio to the local amplitude does not decay;
Ref.~\cite{FeffermanMayboroda2026} controls it uniformly in the number of generations through the estimate recalled in
Section~\ref{sec:props}, by which the coarse flow carries the fine eddy together with the
point at which it is read. In the same spirit,
Eq.~\eqref{eq:frame} holds here only up to a residual of order $A_{j}$: the
exact frame is available only in the multivalued construction, and that exchange ---
exactness of the frame for existence of the field --- is the substance of the paper.

\section*{Data availability}
The code of the construction, the analysis scripts that produce every figure and number in
the paper, the run configurations, the data behind the figures and the supplementary movies
are available at \url{https://github.com/R1967J/kinematic-field}. A single self-contained
generator of the velocity field, which depends on nothing else, is
\nolinkurl{kinematic_field.py} at the top level of that repository.

\begin{acknowledgments}
We thank Leonid Sarieddine for Theorem~\ref{thm:L2-limit} and its proof. This work was supported by a grant from the Simons Foundation (00017375; RJ, SM, CdL, DNS, ME).

This model was developed during a retreat in August 2026 at the CIEM (Centro
Internacional de Encuentros Matem\'aticos\slash International Centre for Mathematical
Meetings), a joint initiative of the University of Cantabria and the City Council of
Castro Urdiales. We are grateful for their hospitality.

AI usage: Denario \cite{Denario2026} was used in the initial stages of this project to explore
ideas autonomously.  Claude Code and Codex were used to assist with coding, computations, and
graphic generation. The ideas, inferences, and results all originate in the (human) brains of
the authors. Claude and ChatGPT were used to check citations, spelling, grammar, and
inconsistencies. The text was written by the authors, who are responsible for the results,
findings, and conclusions of this article.
\end{acknowledgments}

\appendix

\section{Numerical methods}
\label{app:methods}

\paragraph{Field.} Eq.~\eqref{eq:field} is evaluated on an $N\times N$ grid. Each measure
$\mu^{(j,k)}$ is deposited with the bilinear (cloud-in-cell) kernel
$\Lambda(\xv)=(1-|x_{1}|)_{+}(1-|x_{2}|)_{+}$ and the generations are summed by fast
Fourier transform, at cost $O(N^{2}\log N)$. This replaces $\Wv^{j}$ by $\Wv^{j}*\Lambda$,
an equally fixed profile, so Theorem~\ref{thm:band} holds with $\widehat{\Wv^{j}}$
replaced by $\widehat{\Wv^{j}}\widehat\Lambda$. Sampling a centre deposited at a
non-integer position aliases the kernel slightly, but the co-moving arm, whose centres
leave the lattice, returns the same $h$ as the frozen arm to $0.005$ and the same in-band
fractions to $0.022$. On the grid $\|\nabla\!\cdot\!\uv\|_{2}/\|\nabla\uv\|_{2}<10^{-14}$.
The centres are advanced
by a second-order Runge--Kutta step with $\Delta t=1\ll\tau_{J-1}$, with $\uv_{<j}$
interpolated bilinearly. Each generation is rescaled every step, in each arm from its own field, to hold its
r.m.s.\ velocity at its $t=0$ value times $(\sum_{k}g_{j,k}^{2})^{1/2}$ (factor $0.64$--$1.94$). Condition (ND) of Ref.~\cite{FeffermanMayboroda2026} is checked for the $\cos^{2}$
bump by computing its single-eddy structure function in $19$ directions out to $3b_{j}$.

\paragraph{Parameters.} Unless stated otherwise $\alpha=1/3$, $\mathrm{Ku}=1$,
$c_{\mu}=0.3$, $\ell_{0}=N/2$ and $s=\sqrt2$, with generations added while
$\ell_{j}\ge6$ cells: $J=11$ at $N=512$ and $J=13$ at $N=1024$, $\ell_{\min}=8$ in both.
The octave-spaced runs (Fig.~\ref{fig:three}B,C and the comparison of the two arms'
$\chi$) use $s=2$, $J=6$, $N=512$.

\paragraph{Dispersion.} $3000$ pairs per run, $R_{0}=N/64$, $\rho=1.2$, $20$ thresholds,
$t_{\rm end}=2400$; shorter runs are censored and flatten $\lambda$ spuriously. At least $99.7\%$ of the pairs reach every threshold
below $R=150$ at $N=512$. Twenty seeds at $N=512$ and four at $N=1024$; the arms share
seeds and every contrast is paired. Exponents are local logarithmic slopes by centred
differences, errors are standard errors over realisations, and $h$ in
Eq.~\eqref{eq:Don} is the mean over eight diagnostic realisations per arm.

\paragraph{Spontaneous stochasticity.} $\ell_{0}=256$, $T=1000$, $10^{4}$ pairs and four
seeds, $\kappa_{p}=10^{-24}$ to $10^{-2}$; errors are a leave-one-$\kappa_{p}$-out
jackknife on the seed-averaged medians. At short times the scheme returns
$\langle R^{2}\rangle=8\kappa_{p}t$ to $0.1\%$. The backward runs replay the stored
field as $-\uv(\xv,T-t)$ with the same seeds and Brownian streams.

\paragraph{Scalar.} Eq.~\eqref{eq:scalar} is solved pseudospectrally with $2/3$
dealiasing, an exact integrating factor for $\kappa\Delta$ and second-order Runge--Kutta
advection sub-cycled at $\mathrm{CFL}=0.35$; $t_{\rm end}=3000$ for the budget run of
Fig.~\ref{fig:three}B, with $300$ snapshots for the third-order moments, and
$t_{\rm end}=2000$ for the diffusivity scans; averages over the last $40\%$. The damping of
the two lowest shells removes $31\%$ of the production in the budget run and enters
Eq.~\eqref{eq:budget} as $\mathcal S(r)$. The band-forced runs use $N=4096$ ($\ell_{0}=2048$, $s=\sqrt2$, $J=17$,
$\ell_{\min}=8$), $\varepsilon_{f}=9$, $t_{\rm end}=6000$, four seeds and
$\kappa\in\{0.8,1.6\}$ at $\mathrm{CFL}=0.20$; runs sharing a seed share the field and the
forcing history, so the $\kappa$ contrasts are paired, and the window of
Eq.~\eqref{eq:plateauval} is fixed once from $P_{f}$ and the smallest $\chi$.

\section{The measurements in detail}
\label{app:details}

\subsection{Pair dispersion}

\paragraph{Intermittency and the deficit.} With $\xi_{p}$ the exponents of the velocity
structure functions, $\lambda\propto1/\langle T\rangle$ gives $\zeta\simeq\xi_{1}-1$
\cite{BoffettaSokolov2002a} and $S_{2}$ gives $h=\xi_{2}/2$, so intermittency shifts the
deficit by about $\xi_{1}-\xi_{2}/2$, which is non-negative by concavity of $\xi_{p}$ and
vanishes only for a non-intermittent field. The subsection on intermittency below measures
it and tests it by lowering the intermittency of the weights. It describes the frozen control,
but not the co-moving field, whose deficit moves with the intermittency more than three times
as much. We therefore use it only as an order of magnitude. The comparison between the two
arms, which share every amplitude, does not rest on it.

\paragraph{Scaling range and window.} The realised $h(r)$ of the octave-spaced field falls
from $0.73$ at $r=3$ to $0.04$ at $r=100$, identically for $J=6$, $7$ and $8$. The
half-octave plateau widens from $0.73$ decades at $N=512$ to $0.86$ at $N=1024$. The
window $0.078\le R/\ell_{0}\le0.1875$ holds five thresholds over $0.31$ decades and starts
above the release transient at $2R_{0}/\ell_{0}=0.0625$. In exponents the paired contrast
over the window is $\zeta^{\mathrm{co\text{-}moving}}-\zeta^{\mathrm{frozen}}=-0.054\pm0.013$,
which involves no $h$. Table~\ref{tab:res} gives the window at both resolutions.

\begin{table}[t]
\caption{The window $0.078\le R/\ell_{0}\le0.1875$ at both resolutions. $h$ is the
mean over eight diagnostic realisations; the frozen and co-moving arms differ in $h$ by
$+0.002\pm0.005$ and $-0.005\pm0.003$. The $\chi^{2}/\mathrm{dof}$ of $\Dsw(R)$ against
zero uses local slopes on shared trajectories, which are correlated, and measures
pointwise consistency rather than goodness of fit.}
\label{tab:res}
\begin{ruledtabular}
\begin{tabular}{lcc}
& $N=512$, 20 seeds & $N=1024$, 4 seeds\\
\hline
$h$ & $0.299\pm0.013$ & $0.308\pm0.014$\\
$\Dsw$ co-moving & $+0.038\pm0.019$ & $+0.072\pm0.034$\\
\quad $\chi^{2}/\mathrm{dof}$ & $1.53$ & $2.18$\\
$\Dsw$ frozen & $+0.091\pm0.019$ & $+0.137\pm0.047$\\
\quad $\chi^{2}/\mathrm{dof}$ & $9.63$ & $2.82$\\
$\Delta\zeta$, paired & $-0.054\pm0.013$ & $-0.059\pm0.028$\\
\end{tabular}
\end{ruledtabular}
\end{table}

\paragraph{The release transient and the range above the window.} Averaged over
$12\le R\le60$, a band that straddles the end of the release transient, the paired contrast
is $-0.006\pm0.010$. Above the scaling range, where $h$ is no longer constant, the two arms
separate much further, by $\Delta\zeta=-0.13$ and $-0.14$ at the two resolutions
(Fig.~\ref{fig:two}D), but we do not use that range as evidence.

\paragraph{The Kubo number.} The paired experiment at $N=512$ was repeated at
$\mathrm{Ku}=0.25$, $0.5$ and $2$ with eight seeds per arm and everything else unchanged
(Table~\ref{tab:kubo}). One thing moves with $\mathrm{Ku}$: the end of the release transient,
located by the sign change of the paired contrast. For $\mathrm{Ku}\le1$ it sits at
$1.5$--$1.8\,R_{0}$; at $\mathrm{Ku}=2$, where the smallest eddies live twice as long, it sits
at $4.9\,R_{0}=0.15\,\ell_{0}$, and four of the five window thresholds lie inside it, so the
frozen and paired window values at $\mathrm{Ku}=2$ are not quoted. Past the transient, over
$0.16\le R/\ell_{0}\le0.35$, the paired contrast is negative at every $\mathrm{Ku}$, in $43$ of
the $44$ seeds.

\begin{table*}[t]
\caption{The paired experiment at four Kubo numbers ($N=512$, half-octave bands). $\Dsw$ and
$\Delta\zeta$ in the window $0.078\le R/\ell_{0}\le0.1875$, with $h$ from eight diagnostic
realisations per arm; past the release transient, over $0.16\le R/\ell_{0}\le0.35$, where $h$ is
not constant, only $\Delta\zeta$ and the ratio of the frozen to the co-moving separation rate
$\lambda$ are given. At $\mathrm{Ku}=2$ four of the five window thresholds lie inside the release
transient.}
\label{tab:kubo}
\begin{ruledtabular}
\begin{tabular}{lcccccc}
& & \multicolumn{3}{c}{window} & \multicolumn{2}{c}{past the transient}\\
$\mathrm{Ku}$ & seeds & $\Dsw$ co-moving & $\Dsw$ frozen & $\Delta\zeta$ & $\Delta\zeta$ & $\lambda_{\rm fr}/\lambda_{\rm co}$\\
\hline
$0.25$ & 8 & $-0.004\pm0.043$ & $+0.140\pm0.051$ & $-0.142\pm0.036$ & $-0.072\pm0.026$ & $0.958\pm0.013$\\
$0.5$ & 8 & $+0.052\pm0.032$ & $+0.160\pm0.037$ & $-0.110\pm0.014$ & $-0.135\pm0.024$ & $0.828\pm0.016$\\
$1$ & 20 & $+0.038\pm0.019$ & $+0.091\pm0.019$ & $-0.054\pm0.013$ & $-0.207\pm0.020$ & $0.687\pm0.009$\\
$2$ & 8 & $+0.019\pm0.024$ & --- & --- & $-0.222\pm0.055$ & $0.508\pm0.017$\\
\end{tabular}
\end{ruledtabular}
\end{table*}

\paragraph{The eddy arrangement.} In octave-spaced runs ($s=2$, $J=6$, three seeds) the
dispersion exponents of the lattice and of the Poisson set are $-0.7105$ and $-0.7119$,
while switching off the co-motion in the same comparison moves the exponent to $-0.6176$, a
change of $0.093$.

\subsection{Intermittency}

The exponents of Fig.~\ref{fig:four} and of Section~\ref{sec:discussion} come from the frozen control and the field of Definition~\ref{def:main}
at $\mathrm{Ku}=1$ with half-octave bands: eight realisations of each at $N=512$ (seeds $0$ to
$7$) and two at $N=1024$ (seeds $0$ and $1$), with six snapshots per realisation, every $25$
time units up to $t=150$ (\texttt{diagnostics/intermittency.py}, fitted by
\texttt{analysis/intermittency.py}). The increments are taken at fixed time along both grid
axes, so the measurement is Eulerian, and are of three kinds: longitudinal ($u_{x}$ along $x$
and $u_{y}$ along $y$), transverse ($u_{y}$ along $x$ and $u_{x}$ along $y$), and of the full
vector, $|\uv(\xv+r\hat{\mathbf e})-\uv(\xv)|$, whose second moment defines $h$ in
Fig.~\ref{fig:two}A. For each, $S_{p}(r)=\langle|\delta u(r)|^{p}\rangle$ is computed for
$p=0.5$ to $8$ in steps of $0.5$ at $22$ separations spaced geometrically from $2$ to $N/4$
cells, and $\xi_{p}$ is the least-squares slope of $\log S_{p}$ against $\log r$ for the
snapshot-averaged $S_{p}$ of each realisation over the plateau of Fig.~\ref{fig:two}A,
$9\le r\le48$ cells at $N=512$ and $15\le r\le109$ at $N=1024$, above $\ell_{\min}=8$ at
both. Values are means over realisations with their standard errors; with two realisations
the errors at $N=1024$ are only indicative.

A moment is counted as converged when, in every snapshot and at every separation of the
fit, the largest single increment carries less than $1\%$ of the sum that makes up $S_{p}$.
For the vector increment this holds up to $p=3$ at $N=512$ and up to $p=4$ at $N=1024$
($p=5$ in the frozen control).
Beyond, $S_{p}$ is set by a few of the strongest eddies, those in the log-normal tail of
Eq.~\eqref{eq:weights}, the fitted $\xi_{p}$ varies from realisation to realisation and
stops increasing with $p$, and we do not quote it.

Table~\ref{tab:xi} gives the exponents of the vector increment. The longitudinal and
transverse increments give the same picture: $\xi_{p}/p=0.403$, $0.368$, $0.320$
and $0.349$, $0.314$, $0.265$ for $p=1$, $2$, $3$ in the
co-moving field at $N=512$, and over the plateau $\xi_{1}-\xi_{2}/2$ lies between
$0.034$ and $0.039$ for every kind of increment, both arms and both
resolutions. For the vector increment the two arms agree to $0.002$ in $\xi_{p}/p$ for
$p\le3$ at both resolutions.

Over the window $0.078\le R/\ell_{0}\le0.1875$ of Section~\ref{sec:richardson} ($20\le r\le48$ at
$N=512$, $40\le r\le96$ at $N=1024$), $\xi_{1}-\xi_{2}/2$ is $0.028\pm0.002$ in the
co-moving field and $0.027\pm0.001$ in the frozen control at $N=512$, and
$0.025\pm0.002$ and $0.028\pm0.005$ at $N=1024$, the Eulerian estimate of the shift of the
deficit by intermittency (paragraph \emph{Intermittency and the deficit} above); the next subsection tests it. The same fits
give $h=\xi_{2}/2=0.304\pm0.013$ over
the window at $N=512$, against $0.299\pm0.013$ from the diagnostic realisations of
Table~\ref{tab:res}, and $0.308\pm0.017$ at
$N=1024$, against $0.308\pm0.014$. Over the whole plateau the eight realisations give
$h=0.334\pm0.011$ at $N=512$, with a seed-to-seed scatter of
$0.03$, so the single realisation of Fig.~\ref{fig:two}A, with
$h\approx0.31$, is a typical one.

\begin{table*}[t]
\centering
\caption{Exponents of the vector increment, $\langle|\uv(\xv+\mathbf{r})-\uv(\xv)|^{p}\rangle\propto
r^{\xi_{p}}$, over the plateau ($9\le r\le48$ at $N=512$, $15\le r\le109$ at $N=1024$), and
the estimate $\xi_{1}-\xi_{2}/2$ of the shift of the deficit by intermittency, over the plateau and over the window of
Section~\ref{sec:richardson}, with $h=\xi_{2}/2$ there. Errors are standard errors over
realisations. The last row gives the She--L\'ev\^eque exponents of Navier--Stokes
turbulence.}
\label{tab:xi}
\footnotesize
\setlength{\tabcolsep}{3pt}
\begin{tabular}{lccccccc}
\hline
& & & & & \multicolumn{2}{c}{$\xi_{1}-\xi_{2}/2$} & \\
& seeds & $\xi_{1}$ & $\xi_{2}/2$ & $\xi_{3}/3$ & plateau & window & $h$, window\\
\hline
$N=512$, co-moving & 8 & $0.370\pm0.011$ & $0.334\pm0.011$ & $0.287\pm0.012$ & $0.0355\pm0.0011$ & $0.0278\pm0.0023$ & $0.304\pm0.013$\\
$N=512$, frozen & 8 & $0.370\pm0.011$ & $0.335\pm0.012$ & $0.288\pm0.013$ & $0.0353\pm0.0012$ & $0.0274\pm0.0011$ & $0.304\pm0.015$\\
$N=1024$, co-moving & 2 & $0.346\pm0.003$ & $0.310\pm0.003$ & $0.258\pm0.001$ & $0.0359\pm0.0001$ & $0.0245\pm0.0023$ & $0.308\pm0.017$\\
$N=1024$, frozen & 2 & $0.345\pm0.005$ & $0.308\pm0.003$ & $0.257\pm0.001$ & $0.0368\pm0.0024$ & $0.0284\pm0.0048$ & $0.295\pm0.007$\\
She--L\'ev\^eque & & $0.364$ & $0.348$ & $0.333$ & $0.0160$ & & \\
\hline
\end{tabular}
\end{table*}

\paragraph{Weaker intermittency.}

The paired experiment of Section~\ref{sec:richardson} was repeated with $c_{\mu}=0.1$ in Eq.~\eqref{eq:weights},
on the twenty seeds of Table~\ref{tab:res} and with the same protocol: $N=512$, half-octave
bands, $\mathrm{Ku}=1$, $3000$ pairs per arm and realisation, $t_{\rm end}=2400$, and $h$ from
eight diagnostic realisations per arm (\texttt{campaigns/cmu.sh}, \texttt{analysis/cmu.py}).
The weights at different $c_{\mu}$ are made from the same random numbers, so every change with
$c_{\mu}$ is taken seed by seed. The structure functions were measured as in the previous
subsection at $c_{\mu}=0$, $0.1$ and $0.2$, eight realisations each. Table~\ref{tab:cmu} collects the
results. As $c_{\mu}$ falls the moments converge to higher orders and the exponents approach
the non-intermittent line; the drop $\xi_{1}-\xi_{3}/3$ matches She--L\'ev\^eque at
$c_{\mu}\simeq0.12$, by interpolation. At $c_{\mu}=0.1$ the two arms again agree to $0.002$
in $\xi_{p}/p$, and $h$ over the window is unchanged:
$0.301\pm0.011$ (co-moving) and $0.304\pm0.013$ (frozen), against
$0.299\pm0.013$ and $0.301\pm0.015$ at $c_{\mu}=0.3$.

From $c_{\mu}=0.1$ to $c_{\mu}=0.3$ the frozen deficit moves by $+0.015\pm0.015$,
in line with the change of the estimate $\xi_{1}-\xi_{2}/2$ over the window,
$+0.019\pm0.001$. The co-moving deficit moves by
$+0.065\pm0.015$, against $+0.018\pm0.001$ for the
estimate, a difference of $+0.048\pm0.015$
($3.2\sigma$); almost all of it is in
$\zeta$, which changes by $+0.063\pm0.014$, while $h$ changes by
$-0.002\pm0.004$. The runs at $c_{\mu}=0.3$ were made on a different machine from those at
$c_{\mu}=0.1$. Re-running the co-moving seeds $0$ to $4$ at $c_{\mu}=0.3$ on the second machine
changes $\zeta$ over the window by $+0.002\pm0.002$, with a seed-to-seed spread
of $0.005$, so the comparison carries no offset between machines.

\begin{table*}[t]
\centering
\caption{Intermittency of the weights, $c_{\mu}$ in Eq.~\eqref{eq:weights}, against the exponents of the
vector increment over the plateau ($N=512$, co-moving field, eight realisations), the highest
converged order, the estimate $\xi_{1}-\xi_{2}/2$ over the window of Section~\ref{sec:richardson}, and the deficits
of the paired dispersion experiment in that window (twenty seed pairs).}
\label{tab:cmu}
\footnotesize
\setlength{\tabcolsep}{3pt}
\begin{tabular}{lccccccc}
\hline
$c_{\mu}$ & $\xi_{1}-\xi_{3}/3$ & $\xi_{3}$ & $p_{\max}$ & $\xi_{1}-\xi_{2}/2$ & $\Dsw$ co-moving & $\Dsw$ frozen & excess\\
\hline
$0$ & $0.010\pm0.001$ & $1.007\pm0.022$ & $7.5$ & $0.004\pm0.001$ &  &  & \\
$0.1$ & $0.025\pm0.001$ & $0.974\pm0.024$ & $5$ & $0.010\pm0.002$ & $-0.027\pm0.017$ & $+0.075\pm0.018$ & $+0.102\pm0.016$ ($6.4\sigma$)\\
$0.2$ & $0.051\pm0.002$ & $0.925\pm0.028$ & $4$ & $0.019\pm0.002$ &  &  & \\
$0.3$ (paper) & $0.083\pm0.003$ & $0.861\pm0.035$ & $3.5$ & $0.028\pm0.002$ & $+0.038\pm0.019$ & $+0.091\pm0.019$ & $+0.052\pm0.014$ ($3.8\sigma$)\\
She--L\'ev\^eque & $0.031$ & $1$ & & & & & \\
\hline
\end{tabular}
\end{table*}

\paragraph{Clocks.}

The clocks of Fig.~\ref{fig:four}B come from the frozen control and the field of Definition~\ref{def:main} at
$N=512$, $\mathrm{Ku}=1$ and $c_{\mu}=0.3$, four realisations each (\texttt{diagnostics/clocks.py},
\texttt{analysis/clocks.py}). The velocity of generation $j$, $\uv_{<j+1}-\uv_{<j}$, is recorded
every time unit for $300$ units, after a burn-in of $60$, at $4096$ tracers pushed as in the
dispersion runs and at $4096$ fixed points. Its autocorrelation is formed over every time origin,
and $T_{j}$ is the lag at which it first falls to $1/e$. Over $8\le\ell_{j}\le64$,
$T_{j}\propto\ell_{j}^{\,0.679\pm0.005}$ at the fixed points of the frozen field, where
$T_{j}\simeq0.76\,\tau_{j}$ is set by the cross-fade of Eq.~\eqref{eq:window};
$T_{j}\propto\ell_{j}^{\,1.006\pm0.024}$ along its tracers, which are swept through the eddies;
$T_{j}\propto\ell_{j}^{\,1.098\pm0.005}$ at the fixed points of the co-moving field, past
which the eddies are carried; and $T_{j}\propto\ell_{j}^{\,0.807\pm0.047}$ along its tracers,
between $0.73$ and $0.88$ depending on the bands fitted. Paired seed by seed, the ratio of the
co-moving to the frozen clock along tracers grows from $1.10\pm0.06$ at $\ell=128$ to
$1.84\pm0.13$ at $\ell=8$, and at fixed points falls from $0.40\pm0.05$ to
$0.14\pm0.01$. The largest generation, which is moved in neither field, gives
$1.01\pm0.03$ along tracers. Structure functions in time,
$\langle|\uv(\Xv(t+\tau),t+\tau)-\uv(\Xv(t),t)|^{p}\rangle$, have no scaling range at this
resolution: the tracer velocity decorrelates in $28$ to $29$ time units in the
two fields (\texttt{diagnostics/lagrangian.py}, eight realisations each), so we quote no
exponents in time. At short lags the local slope of the second-order function along tracers is
the steeper in the co-moving field, $0.99\pm0.03$ against
$0.83\pm0.02$ at $\tau\simeq7$.

\subsection{The velocity has no cascade}

The statement that the field carries no energy cascade rests on the frozen control and the
field of Definition~\ref{def:main} at $N=512$ with half-octave bands and $\mathrm{Ku}=1$, two
realisations each (seeds $0$ and $1$), with $35$ snapshots per realisation, every $10$ time
units from $t=60$ to $t=400$ (\texttt{diagnostics/cascade.py}).

The skewness of the longitudinal increment,
$\langle\delta u_{L}^{3}\rangle/\langle\delta u_{L}^{2}\rangle^{3/2}$, pooled over the increments
of $u_{x}$ along $x$ and of $u_{y}$ along $y$, lies between $-0.011$ and $+0.029$ at
$r=2$, $4$, $8$, $16$, $32$ and $64$ cells in all four runs. The four-fifths law,
$\langle\delta u_{L}^{3}\rangle=-\tfrac45\varepsilon r$, with
$\langle\delta u_{L}^{2}\rangle\simeq2(\varepsilon r)^{2/3}$, gives $-0.28$ in the inertial range
of three-dimensional turbulence.

With a Gaussian filter of width $\ell$, of transfer function $\exp(-k^{2}\ell^{2}/24)$, the
filtered energy flux $\Pi_{\ell}=-\tau_{ij}\,\partial_{i}\bar u_{j}$, with
$\tau_{ij}=\overline{u_{i}u_{j}}-\bar u_{i}\bar u_{j}$, averages over the domain and the
snapshots to between $-0.046$ and $+0.014$ times its mean magnitude at $\ell=16$, $32$ and
$64$ cells, and is negative on $49.3$ to $52.7\%$ of the area.

In the frozen control these quantities vanish in expectation: the weights of Eq.~\eqref{eq:weights}
are symmetric, so $\uv$ and $-\uv$ have the same law and every odd moment of the velocity is
zero. Its measured values therefore give the sampling scatter, which is set by the few eddies
of the largest generations. The co-moving field lacks the symmetry, since flipping the signs
of the coarse generations changes the paths of the finer centres, but its values fall in the
same range: $-0.002$ to $+0.029$ for the skewness, against $-0.011$ to $+0.028$ in the frozen
control, and $-0.046$ to $+0.002$ for the normalised flux, against $-0.045$ to $+0.014$. The
same symmetry, together with the symmetric cross-fade of Eq.~\eqref{eq:window}, makes the frozen
ensemble invariant under $t\to-t$, $\uv\to-\uv$. In the co-moving field the departure from
that invariance is that each epoch is born on its lattice at the start of its window and not
at its end; the arrangement of the centres barely affects the dispersion
(Section~\ref{sec:richardson}).

\subsection{Spontaneous stochasticity}

Below $\ell_{\min}$ a pair grows exponentially at a rate of order $1/\tau_{\min}$, so it
escapes into the rough range after
$t_{\rm esc}\simeq(\tau_{\min}/2)\ln(\ell_{\min}^{2}/8\kappa_{p}\tau_{\min})$, a time that grows
only logarithmically with $1/\kappa_{p}$; since $\tau_{\min}\propto\ell_{\min}^{1-\alpha}$, the
residual dependence on $\kappa_{p}$ scales with the cutoff as $\ell_{\min}^{1-\alpha}$.
Table~\ref{tab:ladder} gives the exponent for each number of generations. Over
$\kappa_{p}=10^{-7}$ to $10^{-2}$ the median separation at $J=11$ lies between $242$ and $279$
cells, against $\ell_{0}=256$, and the 10th, 25th, 75th and 90th percentiles of $R(T)$,
divided by the median, agree to within $0.025$ at the two ends of that range.

\begin{table}[t]
\caption{Slope of the median separation at time $T$ against the particle diffusivity,
$d\log R_{\rm med}/d\log\kappa_{p}$, as generations are added at fixed $\ell_{0}$, over
$\kappa_{p}=10^{-7}$--$10^{-2}$ and over each cascade's full usable range (decades in
brackets; below it the pairs no longer escape $\ell_{\min}$ by the time $T$);
$\ell_{\min}$ in cells. A field
smooth on the scales the pair explores gives $1/2$; spontaneous stochasticity gives $0$.}
\label{tab:ladder}
\begin{ruledtabular}
\begin{tabular}{rrcc}
$J$ & $\ell_{\min}$ & $10^{-7}$--$10^{-2}$ & full range\\
\hline
$1$ & --- & $+0.4999\pm0.0003$ & $+0.5000$ (22)\\
$6$ & $45$ & $+0.0761\pm0.0149$ & $+0.0761\pm0.0149$ (5)\\
$8$ & $23$ & $+0.0295\pm0.0029$ & $+0.0578\pm0.0091$ (11)\\
$9$ & $16$ & $+0.0198\pm0.0014$ & $+0.0481\pm0.0069$ (15)\\
$10$ & $11$ & $+0.0154\pm0.0003$ & $+0.0417\pm0.0054$ (19)\\
$11$ & $8$ & $+0.0128\pm0.0010$ & $+0.0333\pm0.0038$ (22)\\
\end{tabular}
\end{ruledtabular}
\end{table}

\subsection{The scalar}

\paragraph{The inertial--convective asymptote.} The random source on the lowest wavenumber
shell contributes $P_{f}(r)=4\varepsilon_{f}[1-\rho_{f}(r)]$ to the budget, with $\rho_{f}$
the normalised forcing correlation, in place of the mean-gradient term, and needs no
large-scale damping. Integrating Eq.~\eqref{eq:budget} from the origin then gives an
identity in which every term is measured,
\begin{equation}
\begin{aligned}
\frac{-S_{L}(r)}{r\chi}=1&-\frac{2\kappa}{\chi}\frac{S_{2}'(r)}{r}\\
&-\frac{1}{\chi r^{2}}\!\int_{0}^{r}\!\! r'P_{f}\,\mathrm{d}r'
+\frac{1}{\chi r^{2}}\!\int_{0}^{r}\!\! r'\partial_{t}S_{2}\,\mathrm{d}r' ,
\end{aligned}
\label{eq:asympt}
\end{equation}
with $S_{2}(r)=\langle(\delta\phi)^{2}\rangle$. The first correction is the diffusive
transfer, which is linear in $\kappa$; the other two do not depend on $\kappa$ and are
small. Over the separations between the resolution limit and the forcing scale the raw flux
is $0.825$ at $\kappa=0.8$ and $0.705$ at $\kappa=1.6$, and with the corrections restored
$1.0009\pm0.0026$ and $0.9981\pm0.0026$, which pool to Eq.~\eqref{eq:plateauval}.

\paragraph{The Obukhov--Corrsin exponent.} The band-forced runs give $\beta=0.484$, so
$h+2\beta=1.28$. Under the mean gradient, $\beta$ moves towards the Obukhov--Corrsin value
by $0.115\pm0.018$ when $\ell_{0}/\ell_{\min}$ doubles from $32$ to $64$.

\paragraph{The diffusivity scan.} The straight line $\chi=\chi_{0}-B\kappa$ leaves an r.m.s.\
residual of $0.3\%$, against $1.3\%$ for a power law. The band-forced runs return
$\chi(\kappa/2)/\chi(\kappa)=0.9955\pm0.0043$ because $\langle|\nabla\phi|^{2}\rangle$ doubles,
as the prescribed injection demands.

\section{The centres do not concentrate}
\label{app:centres}

\begin{lemma}\label{lem:count}
Fix an epoch $(j,k)$, let $\Phi=\Phi^{<j}_{(k-1)\tau_{j},t}$ be the flow of $\uv_{<j}$,
and let $B\subset\Tb^{n}$ have finite $(n-1)$-dimensional upper Minkowski content
$\mathcal M^{*}_{n-1}(\partial B)$. Then
\[
\begin{aligned}
\#\{i:\pv^{(j,k)}_{i}(t)\in B\}&=\frac{|B|}{\sigma_{j}^{n}}\\
&\quad+O\Bigl(\frac{\mathrm{Lip}(\Phi^{-1})^{n-1}\mathcal M^{*}_{n-1}(\partial B)}{\sigma_{j}^{n-1}}\Bigr).
\end{aligned}
\]
\end{lemma}
\begin{proof}
The centres in $B$ are the birth lattice points in $\Phi^{-1}(B)$, which has volume $|B|$
because $\nabla\!\cdot\!\uv_{<j}=0$ and boundary $\Phi^{-1}(\partial B)$. Counting
points of spacing $\sigma_{j}$ gives the volume term up to the points in an
$O(\sigma_{j})$-neighbourhood of that boundary.
\end{proof}

The hypothesis is on the Minkowski content and not the perimeter because the count sees
a neighbourhood of the topological boundary, which for a set of finite perimeter can have
positive measure. For the log-normal weights the bound on $\mathrm{Lip}(\Phi^{-1})$ is
measured rather than proved (for bounded weights at small Kubo number it is part of Lemma~2.2 of Ref.~\cite{FeffermanMayboroda2026}):
Gr\"onwall with $\|\nabla\uv_{<j}\|_{\infty}$ is vacuous, since at $N=512$
$\tau_{j}\|\nabla\uv_{<j}\|_{\infty}$ grows from $8$ at $j=1$ to $106$ at $j=9$, but the
r.m.s.\ strain gives $\tau_{j}\langle|\nabla\uv_{<j}|^{2}\rangle^{1/2}=5.5\pm0.4$, flat
from $j=3$ to $j=10$, and $5.2\pm0.2$ at $N=256$, as the self-similarity of
Eq.~\eqref{eq:scales} requires. Individual centres do approach one another: over one
epoch the smallest nearest-neighbour distance falls from $\sigma_{j}$ to
$0.04\,\sigma_{j}$, and the variance of the count in boxes of side $\Delta$, in units of
the mean, is $0.76$, $0.59$, $0.38$ and $0.21$ at $\Delta/\sigma_{j}=1$, $2$, $4$ and $8$,
against $0.99$, $0.98$, $0.93$ and $1.02$ for a Poisson set. Short-range order is lost and
long-range uniformity is kept.

\section{Supplementary movies}
\label{app:movies}

Movies~S1--S4 are single realisations (seed~0) of runs described in the
main text, made with the code of the paper; the numbers quoted in the paper come from the
campaigns of Appendix~\ref{app:methods}. Each is $1920\times1080$ at $25$ frames per second, one frame per
time unit (Movie~S3: one per two). Time is shown in units of $\tau_{\min}=25.4$, the lifetime
$\tau_{j}$ of Eq.~\eqref{eq:scales} of the smallest eddies. Each title below links to the
file in the \nolinkurl{movies/} directory of the repository named under Data availability.

\paragraph{Movie~S1. Push the centres, not the field.} One generation of eddies, of size
$\ell_{j}=\ell_{0}/8=32$ cells, in the octave-spaced field of Fig.~\ref{fig:one}C ($N=512$, $s=2$,
$J=6$), moved three ways from the same random numbers, over the central
$\ell_{0}\times\ell_{0}$ of the domain for $300$ time units. Left: the frozen field. Centre:
the pattern composed with the flow of the coarser generations, the composition route of
Lemma~\ref{lem:dichotomy}. Right: the field of Definition~\ref{def:main}, whose centres are carried by $\uv_{<j}$ and
re-convolved with the fixed profile. Colour: vorticity of that generation; grey arrows:
$\uv_{<j}$. The bars and the strip give the fraction of the generation's energy inside
$[0.7,1.4]\,k_{j}$, the quantity of Fig.~\ref{fig:one}C. Under composition it falls from $0.67$ to
below $0.05$ in about two lifetimes of the smallest eddy and then stays between $0.008$ and
$0.048$; in the frozen field and in Definition~\ref{def:main} it stays between $0.61$ and $0.68$.
\movie{MovieS1_push_the_centres.mp4}

\paragraph{Movie~S2. Who carries the pair.} The frozen control (left) and the co-moving field
(right) of Section~\ref{sec:richardson} at $N=512$ with half-octave bands and $\mathrm{Ku}=1$, from the same
random numbers (seed~0 of the dispersion runs), seen by a fixed camera on the whole periodic
domain for $300$ time units. Colour: vorticity of the generations with $16\le\ell_{j}\le64$
cells, the sizes the pair passes through. Black: one pair released $R_{0}=8$ cells apart,
chosen close to the median separation in both fields at $t=249$, with the last $60$ time units
of its trajectory. In the frozen field the pattern stands still, apart from the renewal of
its eddies, and the pair is swept through it; in the co-moving field the pattern moves with
the flow. Right: the median separation of all $3000$ pairs of the run, which at $t=300$ is
$106$ cells in the co-moving field and $68$ in the frozen one.
\movie{MovieS2_who_carries_the_pair.mp4}

\paragraph{Movie~S3. Born at one point.} $1500$ particles released at one point with
independent Brownian kicks, Eq.~\eqref{eq:spont}, of diffusivity $\kappa_{p}=10^{-2}$,
$10^{-8}$, $10^{-16}$ and $10^{-24}$ (columns), for $1000$ time units. Top: the field of
Definition~\ref{def:main} with eleven half-octave generations ($N=512$, $\ell_{0}=256$), as in
Section~\ref{sec:richardson}; bottom: its largest generation alone, the smooth control, from the same seed.
Colour: vorticity. In the multiscale field each cloud explodes once its particles have
escaped the smooth range below $\ell_{\min}$, after a time that grows like
$\log(1/\kappa_{p})$; in the smooth control the cloud shrinks with $\kappa_{p}$. Right: the
median separation at time $t$ of $2000$ pairs released from coincident points at random
positions, for twelve values of $\kappa_{p}$ from $10^{-2}$ to $10^{-24}$. At $t=1000$
its slope against $\kappa_{p}$ is $0.030$ in the multiscale field and $0.499$ in the smooth
control, against $0.0333\pm0.0038$ and $0.5000$ from the four seeds of Table~\ref{tab:ladder}.
\movie{MovieS3_born_at_one_point.mp4}

\paragraph{Movie~S4. The dissipation does not follow the diffusivity.} The scalar of
Eq.~\eqref{eq:scalar} at $\kappa=6.4$ (left) and $\kappa=1.1$ (right), the end points of the scan of
Fig.~\ref{fig:three}A, in the co-moving field at $N=512$ with half-octave bands, with the same field
and the same source for both. Top: $\phi/\phi_{\rm rms}$; bottom: the local dissipation
$2\kappa|\nabla\phi|^{2}$ on a logarithmic scale common to both; central
$\ell_{0}\times\ell_{0}$ of the domain. Right: $\chi(t)$, after a spin-up of $600$ time
units (shaded), for the $300$ time units shown. Over that interval $\chi$ averages $16.0$ at
$\kappa=6.4$ and $19.7$ at $\kappa=1.1$, a ratio of $1.23$ for a factor $5.8$ in $\kappa$;
without an anomaly the ratio would be $0.17$. This single realisation corresponds to
$d\log\chi/d\log\kappa=-0.12$; Eq.~\eqref{eq:anomaly} averages six realisations and four
diffusivities. The movie integrates the scheme of Appendix~\ref{app:methods} with real transforms.
\movie{MovieS4_dissipation_vs_diffusivity.mp4}

\end{document}